\documentclass[preprint,12pt,authoryear]{elsarticle}

\usepackage{amssymb}
\usepackage{graphicx}	
\usepackage{amsmath}	
\usepackage{rotating}
\usepackage{pdflscape}
\usepackage{longtable}
\usepackage{booktabs}
\usepackage{afterpage}
\usepackage[T1]{fontenc}

\journal{Machine Learning for Small Bodies in the Solar System}

\begin{document}

\begin{frontmatter}



\title{Identification and Localization of Cometary Activity in Solar System Objects with
Machine Learning}


\author{Bryce T. Bolin}

\address{organization={Goddard Space Flight Center},
            addressline={8800 Greenbelt Road}, 
            city={Greenbelt},
            postcode={20771}, 
            state={MD},
            country={USA}}
            
\author{Michael W. Coughlin}

\address{organization={School of Physics and Astronomy, University of Minnesota},
            addressline={116 Church St SE}, 
            city={Twin Cities},
            postcode={55455}, 
            state={MN},
            country={USA}}

\begin{abstract}
In this chapter, we will discuss the use of Machine Learning methods for the identification and localization of cometary activity for Solar System objects in ground and in space-based wide-field all-sky surveys. We will begin the chapter by discussing the challenges of identifying known and unknown active, extended Solar System objects in the presence of stellar-type sources and the application of classical pre-ML identification techniques and their limitations. We will then transition to the discussion of implementing ML techniques to address the challenge of extended object identification. We will finish with prospective future methods and the application to future surveys such as the Vera C. Rubin Observatory.
\end{abstract}




\end{frontmatter}


\section{Introduction to the identification of cometary activity in Solar System objects}
\label{sec: mlc.intro}

Current and next-generation all-sky surveys provide unprecedented opportunities for the discovery of solar system asteroids and comets \citep[][]{Jedicke2016, Bauer2022cometsurveys}. Contemporary asteroid surveys employ a variety of algorithms and pipelines to detect and identify moving objects in their image data such as the association of two or more detections taken in a sequence called ``tracklets''  \citep[e.g.][]{Jedicke2015} or through the shifting and stacking of multiple images together \citep[e.g.][]{Whidden2019,Bolin2020CD3}. This involves taking multiple images to detect an object more than once within a short enough time for the detections to be linked. The actual linkage of detection can include decision trees or the connection of several detections together within similar co-moving velocities \citep[][]{Kubica2005,Masci2019}.

Multiple detection linking algorithms can link the detections of both point-source, asteroidal detections and those that are extended, as for comets. Even for comets, whose detections have a large, extended appearance, multiple detections can be linked as for point sources if the measurement of the comet's position relative to the moving frame of the object is consistent from detection to detection \citep[][]{Denneau2013}. However, in some survey imaging pipelines, extended objects, defined as having a width more expansive than the measured width for known point sources, can be flagged as potential outliers and removed from further processing \citep[][]{Duev2019}. The similarities between extended objects and common telescope/detector artifacts can cause the source detection pipeline to fail; therefore, the flagging and removal of extended objects can be motivated by the need to keep the false positives at manageable levels for vetting by human eyes. As a result, the detection and discovery of comets by all-sky surveys can depend not only on the ability of the algorithms to link the detection of comets from multiple images but also on the base level of extraction of detections on a per-image basis.

The ability to detect comets can vary from survey to survey due to the ability to detect extended objects. For example, the image processing pipeline for the ZTF survey rejects extended objects due to the contamination of artifacts \citep[][]{Masci2019}. Therefore, the detection of comets is significantly limited in surveys such as ZTF using the standard data pipeline. Furthermore, the ability to link comet detections can be substantially different than the detection and linking of asteroidal or point-source-like detections. The detection and discovery of comets may require their detection to be non-extended to avoid being flagged as an artifact. Examples of such cases are where the appearance of a comet is non-extended, either by its extent not being detectable given the seeing conditions at the observing location of the survey \citep[typically 1-2 arcseconds][]{Jedicke2015} or by being discovered as a bare nucleus before their activity starts \citep[e.g.,][]{Cheng2022A}. In this review chapter, we will focus on identifying comets as truly extended sources, i.e., that have morphological features that are recognized as extended at the resolution of the survey. We will describe classical methods of identifying comets in astronomical survey data as well as by follow-up data. We will also describe recent advances in machine learning techniques applied to observations of comets that assist in their recognition as extended cometary sources versus being asteroidal, focusing on advancements in Deep Learning. Lastly, we will describe techniques that are being mastered for near-future use for the detection and discovery of comets in astronomical surveys such as the forthcoming Vera C. Rubin Observatory's Legacy Survey of Space and Time \citep[LSST, ][]{Schwamb2023ApJS}.

\section{Review of classical comet detection methods and strategies}

We will start this review with an overview of contemporary methods for finding comets in astronomical surveys and follow-up data. We will start this section with a brief discussion of ''classical'' comet detection methods, i.e., those not based on machine learning or Deep Learning methods. We refer to these methods as classical since they were, in principle, accessible before the wide-scale adoption of computational methods in astronomy, i.e., when the discovery of comets occurred by recognition of an extended object by the human eye \citep[][]{Festou1993}. However, for this review, we set a lower bound to the time covered by classical comet identification methods to the early to late 1990s when consumer-grade charge-coupled devices (CCDs) became widely accessible for use in astronomical surveys. We will transition from classical methods to methods based on characterizing individual detections to data science and machine learning-driven methods, which became more prevalent in the 2010s.

\subsection{Serendipitous identification of cometary activity of unknown objects}

The first discovery of a comet as an unknown object with an automated asteroid detection pipeline was made in 1992 of C/1992 J1 (Spacewatch) by the Spacewatch survey \citep[][]{Scotti1992}. Spacewatch was one of the first surveys to use an automated pipeline for the detection of asteroids in CCD observations, while the telescope was used in drift scan mode \citep[][]{Jedicke2015}. Several asteroid surveys using CCD technology began after Spacewatch such as the Catalina Sky Survey, Lincoln Near-Earth Asteroid Research, Panoramic Survey Telescope and Rapid Response System  (Pan-STARRS) \citep[][]{Chambers2016}, the Near-Earth Object Wide-field Infrared Survey Explorer (NEOWISE) \citep[][]{Mainzer2011a}, Asteroid Terrestrial-impact Last Alert System (ATLAS) \citep[][]{Tonry2018}, and the Zwicky Transient Facility (ZTF) \citep[][]{Bellm2019} \citep[for a review of asteroid surveys, please see][]{Jedicke2015}. These ground and space-based surveys discovered numerous comets, $\sim$1000 by the end of the 2020s \citep[][]{Bauer2022cometsurveys}. The SOHO and STEREO observatories also discovered several thousand comets as they passed near the Sun in the field of view of these spacecraft \citep[for an overview, please see][]{Battams2017}. However, for this review, we will focus on comet discoveries made by asteroid-dedicated surveys.

\subsection{Identification of cometary activity in known objects}

As described above, the discovery of cometary activity can be decoupled from the initial recognition of a solar system object, i.e., it will be recognized as a candidate asteroid detection agnostic of its activity. A known candidate orbit is the main product of a series of observations of an asteroid candidate. Even if the detection of a candidate comet's activity is not known, the orbit resulting from the observations can provide clues as to its cometary nature. A clue to the cometary nature of a comet candidate can be deduced from the known orbit such as having a Tisserand parameter significantly less than 3 \citep[for a review of the Tisserand parameter and its significance to comets, please see][]{Fraser2022review}.  In addition, having an eccentricity of $\gtrsim$1 or retrograde inclination may suggest that an asteroidal object has a cometary origin versus an asteroidal one. This is because most observed asteroids, such as those in the Main Belt, have an eccentricity less than 0.3 and an inclination less than 17 degrees \citep[e.g.,][]{Bolin2017,Bolin2017b}

Comets which from their orbital properties provide clues of cometary origins may encourage follow-observations that are deeper in limiting magnitude and sensitivity which may result in the detection of cometary activity. A recent example is in the discovery of C/2022 E (ZTF) \citep[][]{Bolin2022MPECE3} which, while possessing evident activity in subsequent observations, had an asteroidal appearance in the discovery images taken by ZTF. Subsequent observations taken by amateur astronomers specializing in the follow-up of comets, as well as in subsequent observations by ZTF, showed evidence of cometary activity in the form of a tail and coma \citep[][, Fig.~1]{Sato2022E3,Bolin2024E3}. Clues to its cometary origin, which inspired follow-up observations, came from its eccentricity of $\sim$1 and retrograde inclination of $\sim$109 degrees, typical for long-period comets and distinct from the population of asteroids \citep[][]{Jedicke2002,Gladman2008,Nesvorny2023NEO}.

\begin{figure}\centering
\includegraphics[width=0.7\linewidth]{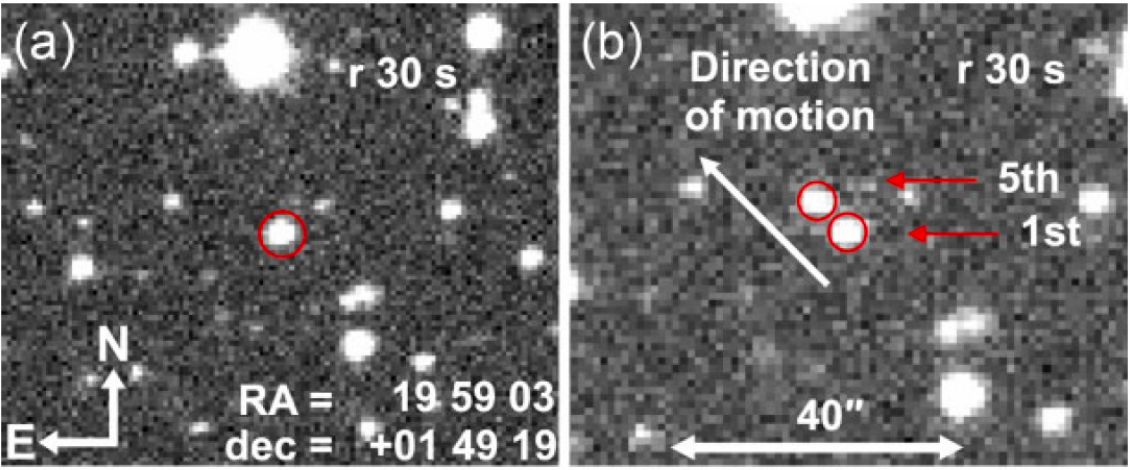}
\includegraphics[width=0.4\linewidth]{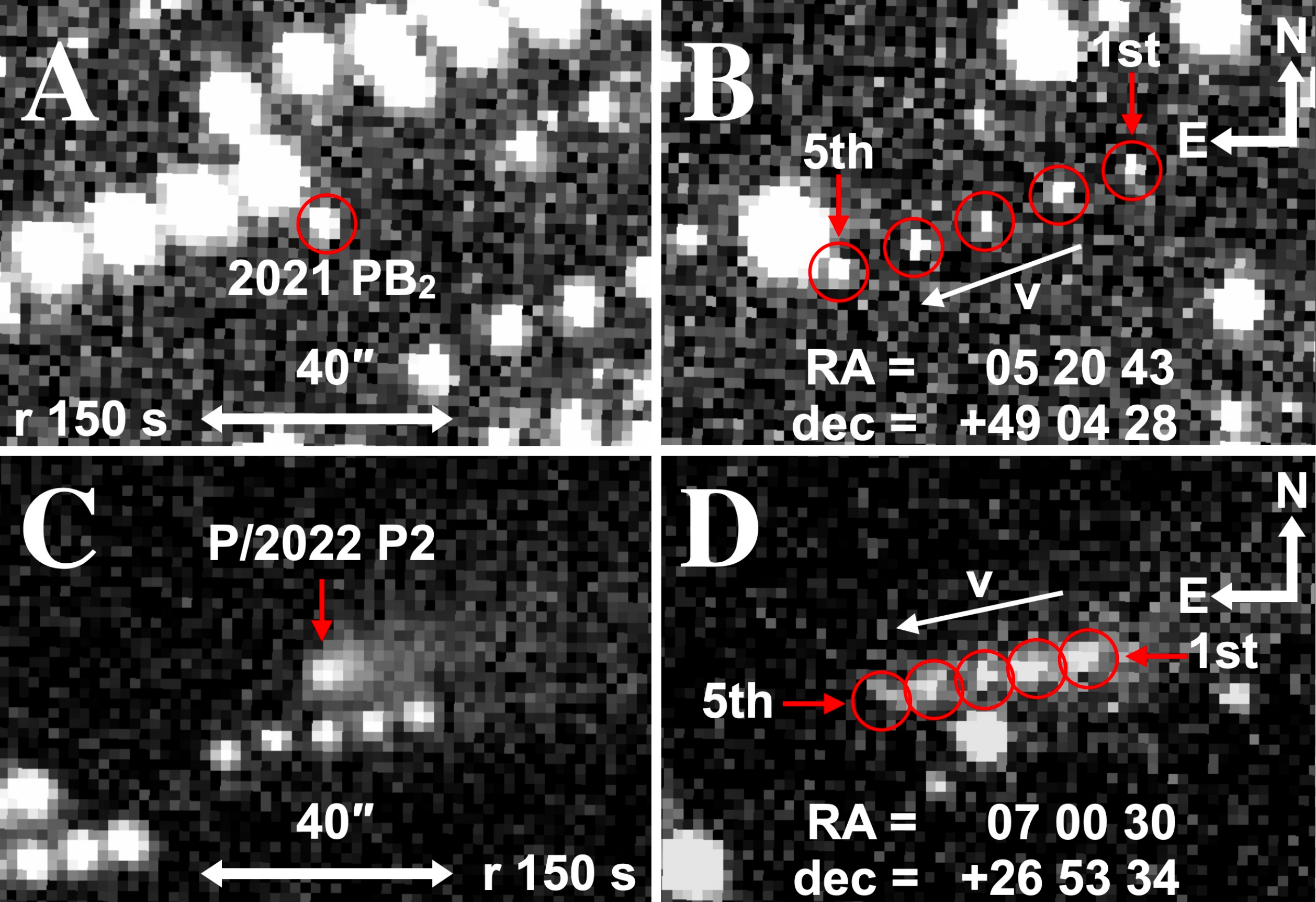} 
\caption{Discovery images of C/2022 E3 (ZTF) taken by ZTF in 2022 March. The top panel shows one of the individual detections of the comet (taken in a series of 5). The top right panel shows a composite stack of the first and fifth images stacked on the background stars. The bottom panel shows a composite stack of all five images co-added together. The exposure time, filter, and cardinal directions are indicated. Adapted from Figure 1 of \citet[][]{Bolin2024E3} and reproduced with the permission of the authors and MNRAS. The bottom panel shows a co-added composite stack of all five 30 s exposures of images of comet P/2022 P2 (ZTF) taken by ZTF in 2022 August \citep[][]{Bolin2024TS}.}
\end{figure}

\section{Review of cometary detections methods based on measurements of the individual detections}

Next, we discuss detecting cometary activity by inspecting objects on an individual image basis. Previous techniques relied on identifying comets following the linkage of multiple detections of objects taken in multiple images. The next set of techniques relies on the previous knowledge of the location of a known object in CCD images and inspects each detection individually. Thus, the detection of cometary activity can be done with individual images instead of grouped images.

\subsection{Identification of comets through spread function parameter analysis}

Extended objects in astronomical images, whether in-ground or space-based observations, are defined by their extent relative to the typical width of unresolved detections. In images taken by single-dish telescopes, the width of the point spread function (PSF), the detection of a point or unresolved source, is set by the diffraction limits of the telescope optics, directly proportional to the observed wavelength and inversely proportional to the diameter of the telescope \citep[][]{Born1999}. In CCD detections, the PSF can be super-sampled, that is, when the scale of the individual pixels is less than the critical sample rate, and sub-sampled if the pixel scale is larger than the critical sampled rate \citep[][]{Howell2000}. 

Detecting extended cometary features such as a tail or coma depends on their angular size, which is significantly larger than the imager's resolution. Even when observing with a telescope that enables resolving objects at a diffraction limit smaller than the angular scale of an extended comet, the extendedness may not be resolved if the effective resolution of the images is too coarse due to pixel size. Additionally, the resolution is often limited by the quality of atmospheric seeing when observing from the ground (typically $\sim$1 arcsecond at most professional observing sites) rather than being limited by the inherent diffraction limit of the telescope's optics.

Several attempts have been made to identify comets by quantification of the quality and extendedness of their PSFs. The quality of a PSF is roughly indicated by the reliability of PSF shape measurements made of asteroid detections by imaging pipelines \citep[e.g., as for Pan-STARRS, see ][]{Chambers2016}. For the identification of comet candidates, the measured size of the PDF is compared to the expected PSF size for a non-resolved, point-like source. Active comet candidates are more likely to have PSFs that are significantly more extended than compared to stellar-like sources (top panel of Fig.~2). Studies of the detection of comets at Palomar observatory and with Pan-STARRS found a significant correlation between parameters describing the extent and quality of PSF for comets when compared to asteroidal detections (see Fig.~2) \citep[][]{Waszczak2013,Hsieh2015}. This method resulted in the discovery of numerous comets with Pan-STARRS, including several Main Belt comets/Active asteroids \citep[e.g.,][]{Bolin2013,Hill2013b}.

\begin{figure}\centering
\includegraphics[width=0.7\linewidth]{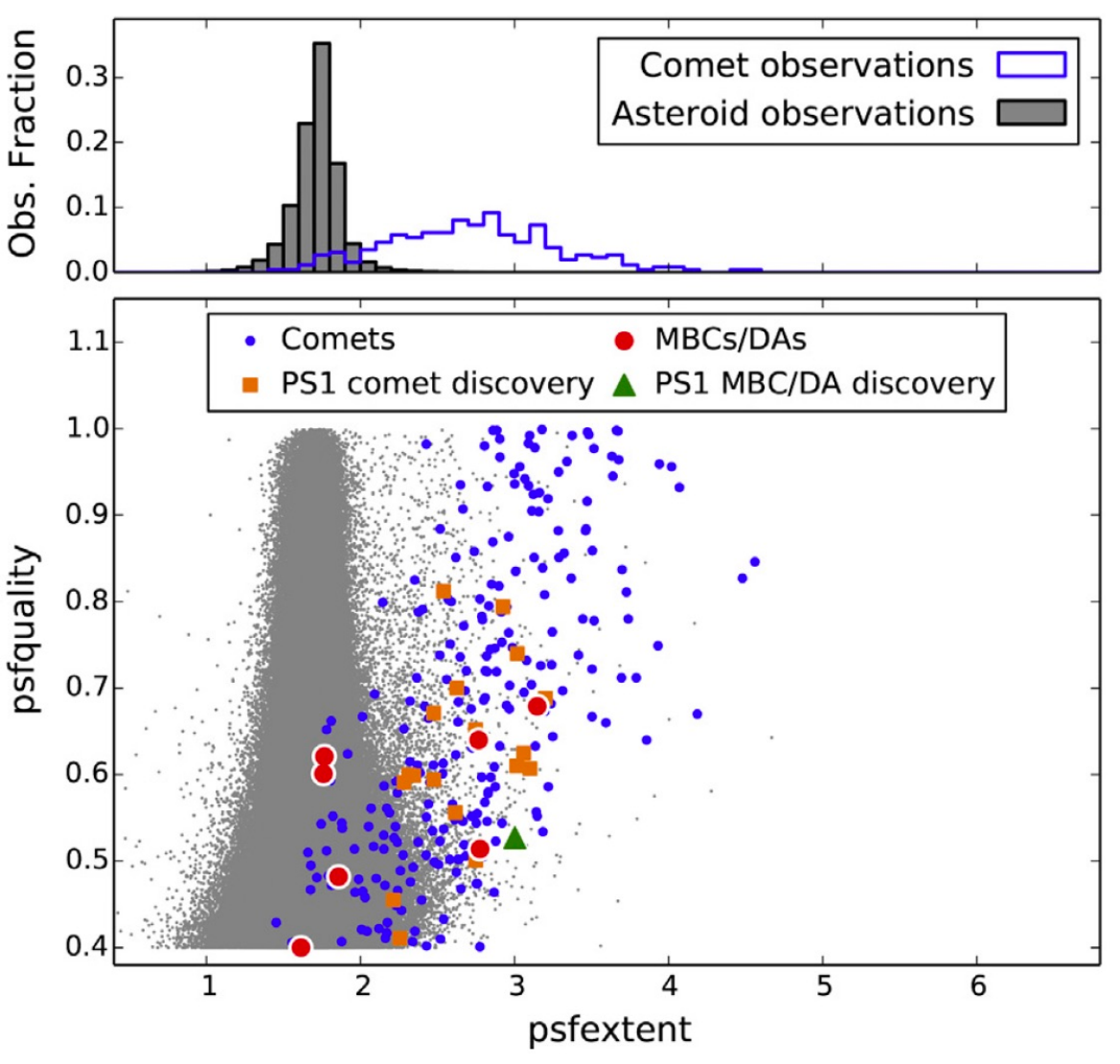} 
\caption{Comparision of the PSF extent, psfextent, and PSF quality, psfquality,  properties of asteroid and comet detections from Pan-STARRS taken before 2013 July. psfextent refers to the extendedness of detection PSFs measured as the full width at half maximum of the PSF in arcseconds compared to the expected PSF of point sources.  Higher values of psfextent correspond to sources being more extended. psfquality refers to the reliability of PSF measurements with a higher number being more favorable. Top panel: 1-D histograms of PSF extent of asteroid detections (grey) and comet detections (blue).  Bottom panel: 2-D distribution of psfextent vs psf quality. The psfextent and psfquality for asteroids is in grey, comets in blue/orange, and Main Belt comets/active asteroids in red and green. Adapted from Figure 5 of \citet[][]{Hsieh2015} and reproduced with the permission of the authors and Icarus. We refer readers to the asteroid and comet detection gallery seen in Fig.~18 of \citet[][]{Denneau2013} for examples of asteroidal and extended comet detections.}
\end{figure}

\subsection{Identification of cometary activity by detection of comae and azimuthal variations}

While the PSF extendedness method provides a straightforward and intuitive method for flagging comet candidates by their candidate extendedness when compared to reference point-source PSFs, this comparison is not based on the observed attributes of comets. Comets that can have azimuthal asymmetry due to the presence of tails or a central extended compact coma that may not be easily represented by a PSF \citep[e.g.,][]{Bolin2020HST,Bolin2021LD2}. Improvements to searching for cometary signatures in candidate comet detections can be made by searching for azimuthal asymmetry and the presence of a coma with special aperture functions designed to detect these properties.

\begin{figure}\centering
\includegraphics[width=0.7\linewidth]{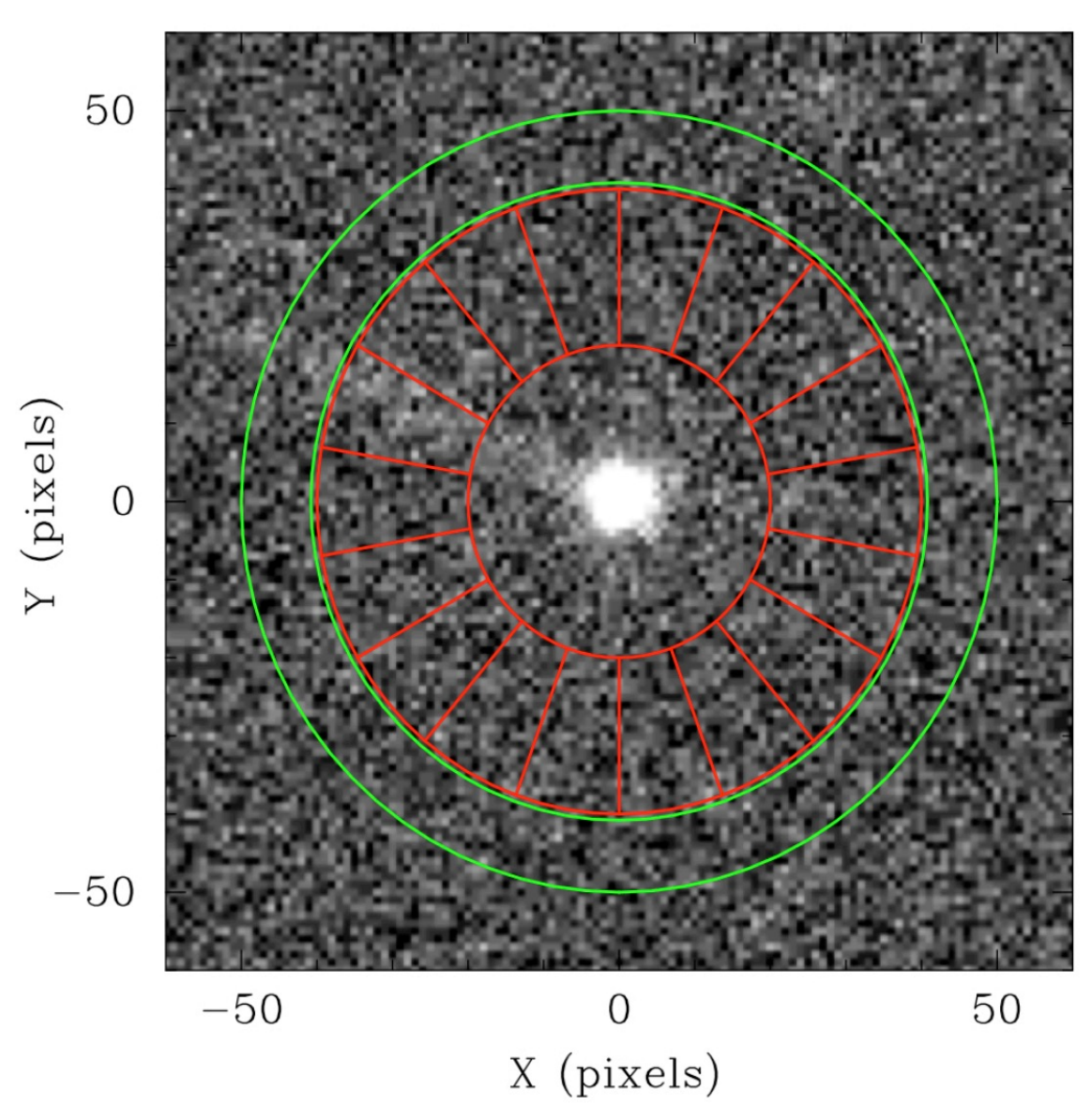} 
\caption{Azimuth scheme for the detection of tails of comets in individual CCD images. The azimuth ring is divided into 18 sections. The section at 10 o'clock encompassing the tail of the detected comet, 133P, measures a higher degree of flux compared to the other sections. The radius of the azimuthal sections was chosen to avoid background stars. Adapted from Figure 2 of \citet[][]{Sonnett2011} and reproduced with the permission of the authors and Icarus.}
\label{fig:azimuth}
\end{figure}

\citet[][]{Sonnett2011} were the first to apply the azimuthal asymmetry and coma search technique to $\sim$1000 asteroids detected by the Thousand Asteroid Lightcurve Survey taken with the Canada France Hawaii Telescope \citep[][]{Masiero2009}. They used a scheme shown in Fig.~\ref{fig:azimuth} that divides an annulus outside the extent of the central coma into 18 sections. The section containing part(s) of a candidate comet's tail will be measured to have a higher flux than the other surrounding sections. \citet[][]{Sonnett2011} used this technique to set a 90$\%$ confidence upper limit to the number of Main Belt comets larger than 150\,m to be $\sim$400,000.

\citet[][]{Chandler2021} and \citet[][]{Ferellec2023} used similar azimuthal and coma detection techniques applied to comet candidates observed by the Blanco 4.0-m/Dark Energy Camera and the Isaac Newton Telescope/Wide Field Camera. \citet[][]{Chandler2021} identified activity in asteroid (248370) 2005 QN$_{173}$ using this technique finding evidence of its recurrent activity throughout its orbit. \citet[][]{Ferellec2023} detected an elevated coma and azimuthal asymmetric flux for the Belt asteroid  (279870) 2001 NL$_{19}$ and set an upper limit to the occurrence rate of Main Belt comets of 1:500. 

\subsection{Detection of cometary activity through sporadic, significant deviations from phase curve brightness}

The brightness of asteroids or bare, inactive comet nuclei follow a steady phase function determined by heliocentric, geocentric, and phase angle parameters \citep[][]{Bowell1988, Muinonen2010}. The phase function predicts an asteroid's brightness will increase as it gets closer to the Earth, the Sun, and for a smaller phase angle. In general, this phase function is described by a smooth change in brightness for distant asteroids as their viewing geometry changes due to the motion of their orbit and their evolving view from the Earth.

\begin{figure}\centering
\includegraphics[width=0.7\linewidth]{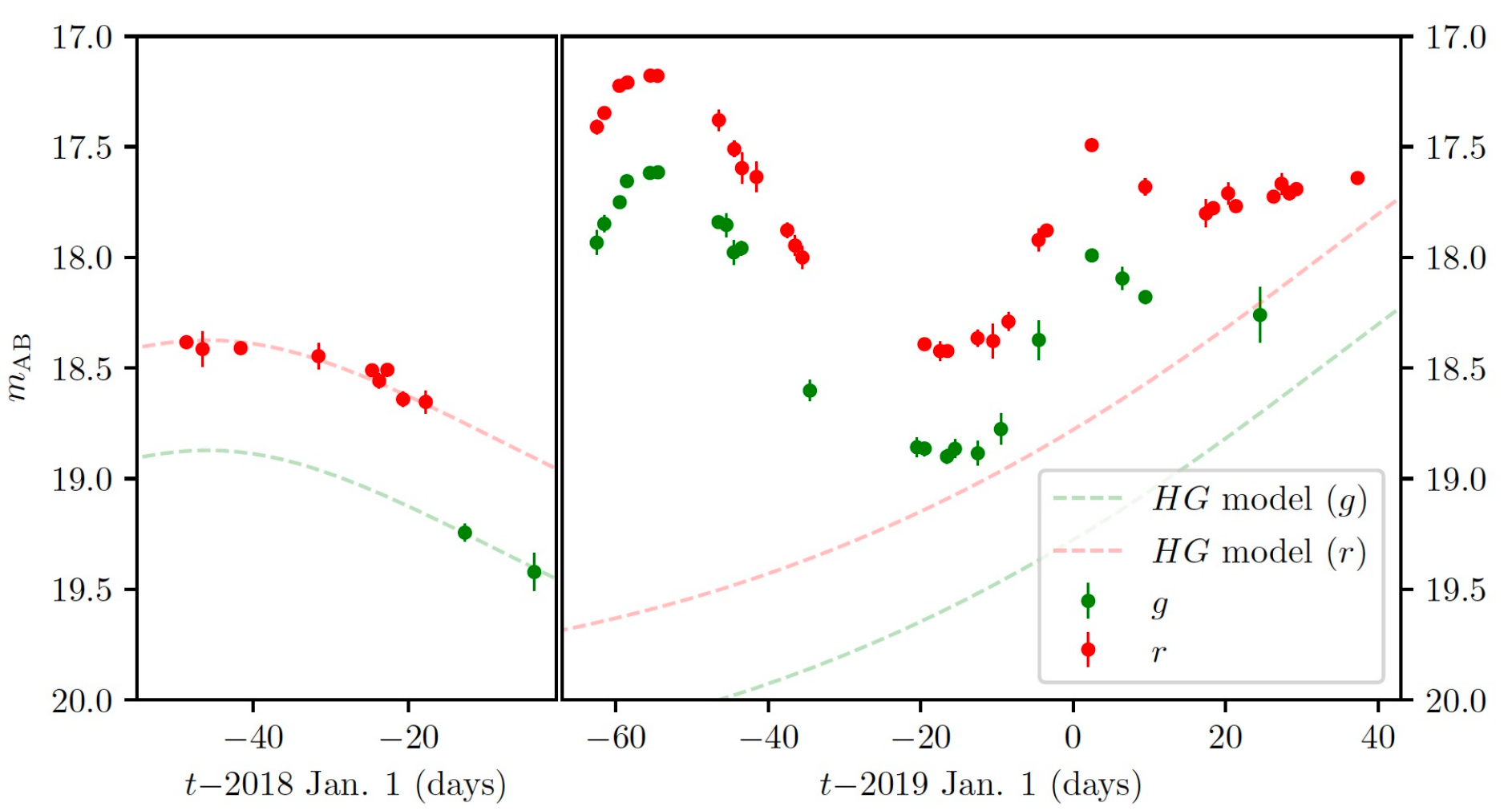} 
\caption{Secular brightness evolution of active asteroid (6478) Gault in 2017-2019. The left and right panels show the evolution in the comet's brightness (red and green dots) compared to brightness models for a bare nucleus (red and green dashed lines). The brightness evolution of Gault in 2017 follows a typical pattern for a bare nucleus. The brightness pattern significantly deviates and brightens compared to a bare nucleus in 2018-2019. Adapted from Figure 2 of \citet[][]{Ye2019Gault} and reproduced with the permission of the authors and AAS Journals.}
\label{fig:Gault}
\end{figure}

Monitoring the change in brightness of comet candidates with respect to the secular phase function assuming a bare nucleus provides a test of the presence of sudden outbursts or sporadic activity \citep[][]{Kelley2019Zchecker}. The brightness of a bare nucleus follows a smoothly varying function describing the increase in an object's brightness as a function of phase angle, heliocentric distance and geocentric distance \citep[][]{Bowell1988, Muinonen2010}. This technique has been applied to the detection of activity in active asteroid (6478) Gault as seen in Fig.~\ref{fig:Gault} where the brightness of Gault increased by several magnitudes in late 2018/early 2019 ZTF data compared to its phase function \citep[][]{Ye2019Gault, Purdum2021}. A similar technique was applied to data from ATLAS for the detection of activity in Centaur comets \citep[]{Dobson2023}.

\section{Review of cometary detection methods based on machine learning methods}

The next set of methods is based on the growing set of techniques related to Machine Learning (ML). Carruba et al. have already described many of the core algorithms behind these techniques in Chapter 1 of this volume. We will, therefore, provide an overview of ML-based comet discovery techniques, highlighting where many of these algorithms have been implemented to find comets. Methods such as crowdsourcing used to find comets, while formidable, are outside the scope of this chapter.

\subsection{Identification of cometary objects with convolutional neural networks}

Fully-connected neural networks (NNs) have been used in machine vision applications such as in image and facial recognition applications \citep[e.g.,][]{Bengio2009}. Applications of NNs adjacent to the recognition of comets include the Morpheus network, a network designed to identify extragalactic sources in \textit{Hubble Space Telescope} image data \citep[][]{Hausen2020}. The training of NNs usually involves training with a gradient descent-like algorithm to converge on minimum-cost function trained model \citep[e.g.,][]{Kolen2001}. Fully connected networks are computationally expensive to train and run since they require weights for every hidden layer. However, convolutional neural networks (CNNs) provide a possible workaround by downscaling the first connected layer through convolution and max pooling (see Section 3 of Chapter 1 of this volume for details). This provides an order of magnitude fewer parameters to train and a corresponding considerable speed up in training and reduction in run time while maintaining accurate results compared to fully connected networks.

\subsubsection{training of comet identification convolutional neural network models}

Previous comet-recognition methods using CNNs relied on a two-stage process combining classical moving object detection methods with CNNs \citep[][]{Rabeendran2021}. The Tails network by \citet[][]{Duev2021} is the first example to use a CNN trained on comet data to identify comets on a per-image basis.  The architecture of the Tails network is provided below in Fig.~5. The training set consisted of $\sim$3,000 image examples of comets with identifiable cometary morphology; an example can be found in Fig.~6. Each training example consisted of a science image, a reference image consisting of deep stacks of previous images, and an image consisting of the difference between the two. Approximately 20,000 negative examples of point source-like detections, internal reflection artifacts, hot pixels, and diffraction spikes were used as negative examples. The training set was refined with several instances of active learning. Active learning is an interactive form of machine learning in which the user manually labels data to improve the accuracy of a classifier \citep[][]{Fang2017AL}. In the development of Tails, active learning was used in cases near the network's decision threshold that were visually inspected, then re-classified and re-added to the revised training set.

\begin{figure}\centering
\includegraphics[width=0.9\linewidth]{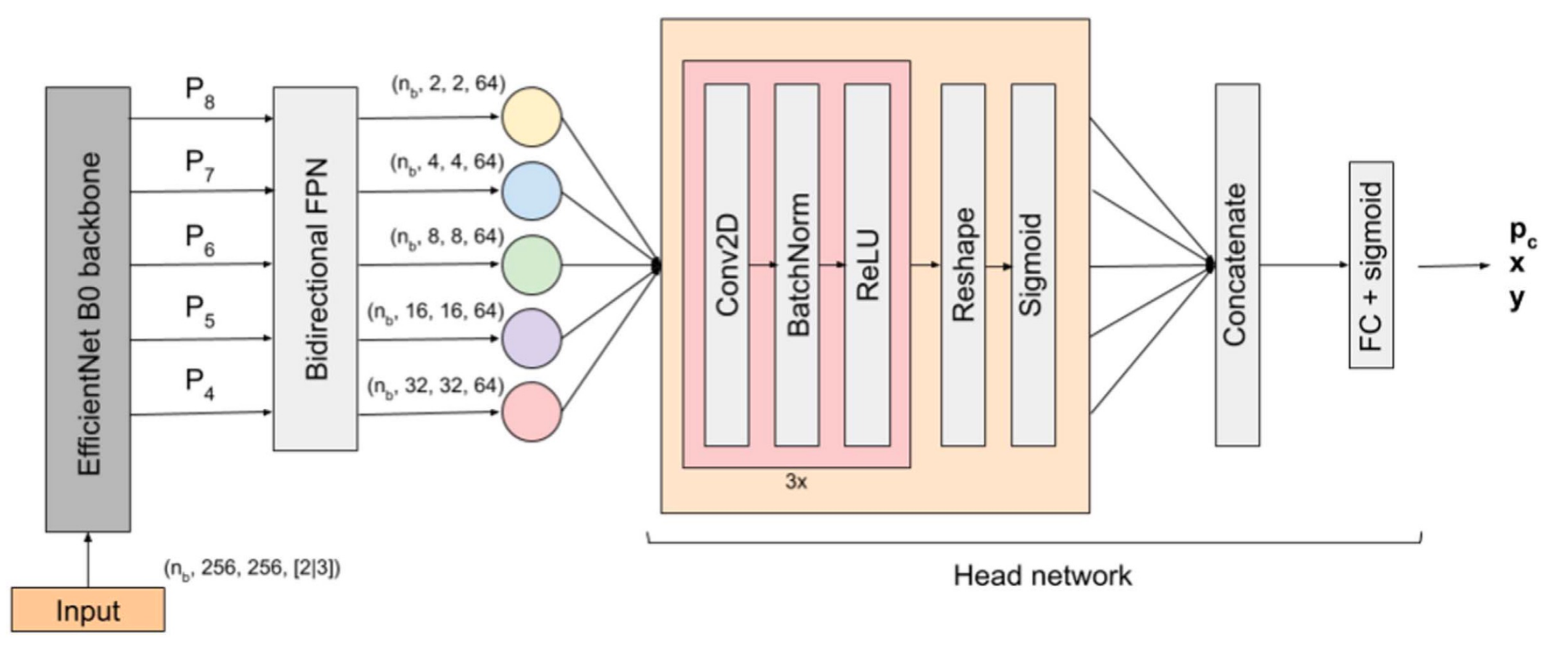} 
\caption{Illustration of the Tails network architecture. Tails uses a combination of bidirectional feature pyramid networks (BiFPN), which are fed by the last five layers of an EfficientDet network. The output of the BiFPN is fed into a head network consisting of several convolutional layers and a fully connected layer and activation function to derive a comet probability score between 0 and 1. Adapted from Fig.~3 of \citep[][]{Duev2021} and reproduced with the permission of the authors and AAS Journals.}
\label{fig:tails}
\end{figure}

\begin{figure}\centering
\includegraphics[width=0.8\linewidth]{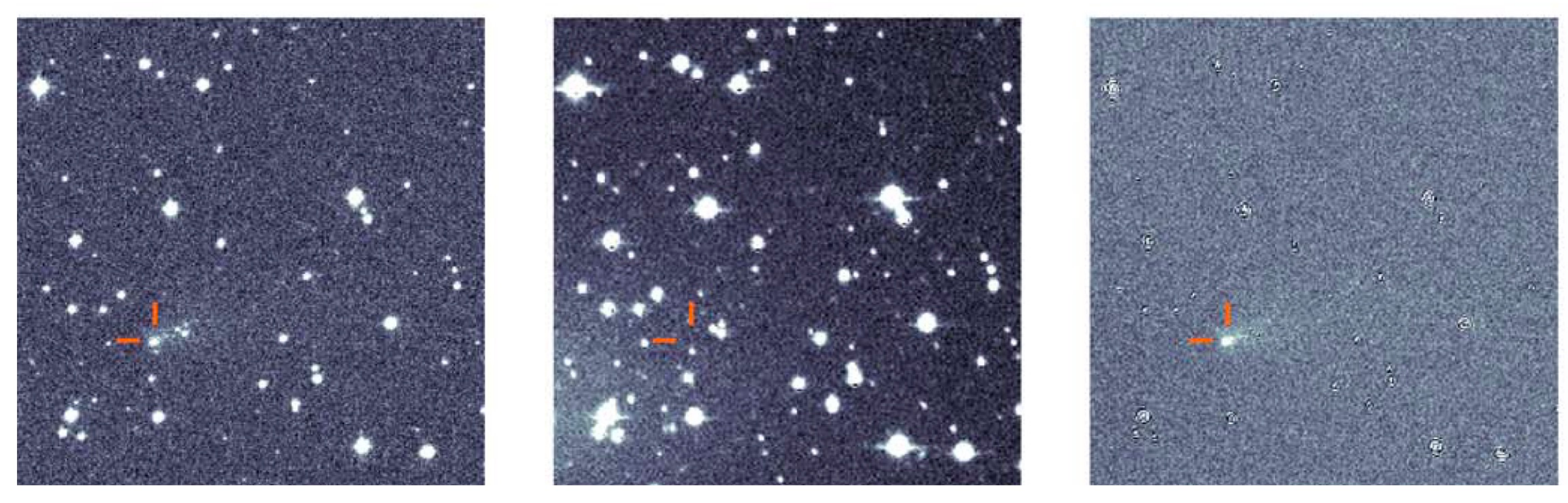} 
\caption{Example of detection of comet C/2019 D1 (Flewelling) used to train the Tails CNN. The left panel shows the science images, the central panel shows the reference panel and the right panel shows the difference between the science and reference image. Adapted from Fig.~2 of \citep[][]{Duev2021} and reproduced with the permission of the authors and AAS Journals.}
\label{fig:tails_example}
\end{figure}

\subsubsection{Performance of convolutional neural networks to identify comets}

\begin{figure}\centering
\includegraphics[width=0.8\linewidth]{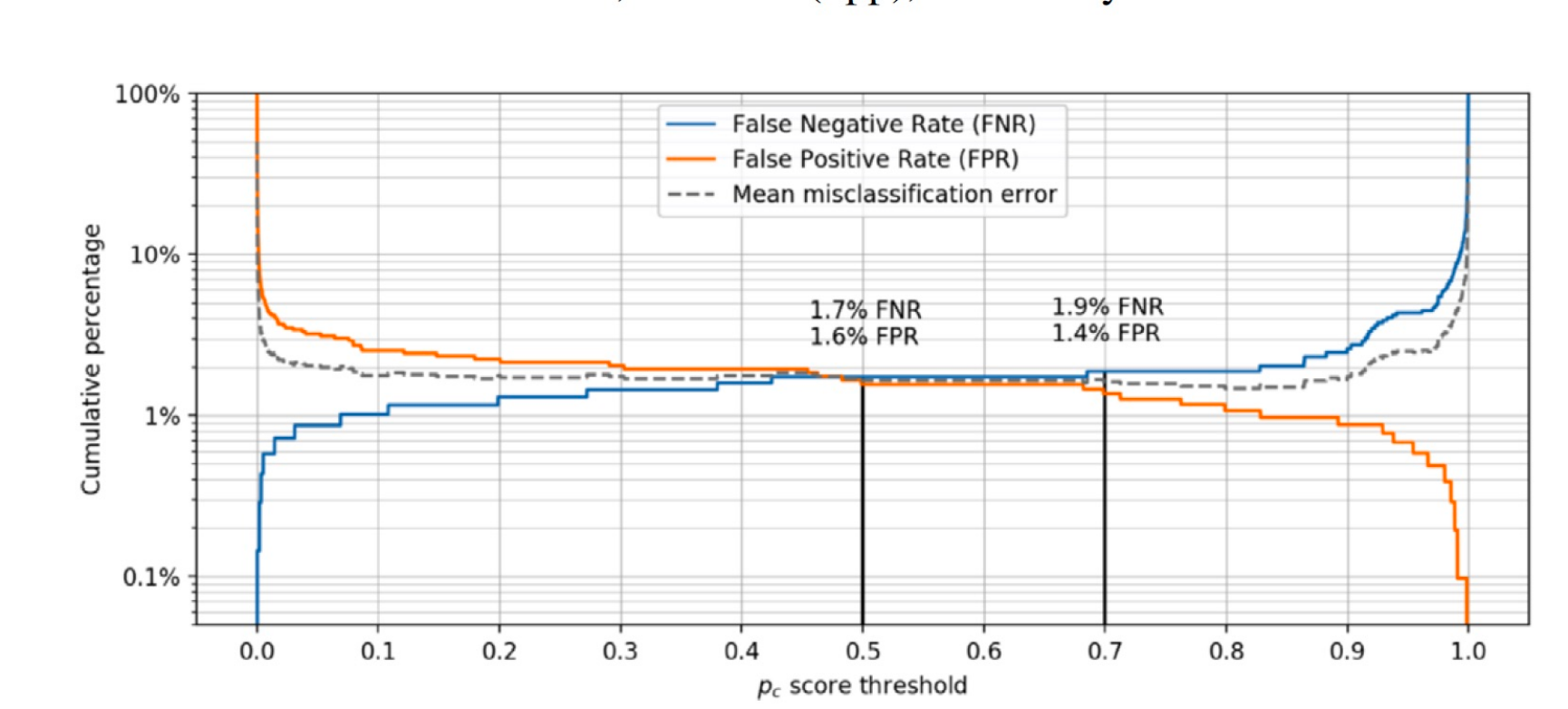} 
\caption{False positive and false negative rates as a function of comet probability score. The False positive and false negative equalize at $\sim$1.7 percent for a comet probability score of 0.5 and bifurcate into a false positive rate of $\sim$1.4 percent and a false negative rate of $\sim$1.9 percent for a score of 0.7. Adapted from Fig.~4 of \citep[][]{Duev2021} and reproduced with the permission of the authors and AAS Journals.}
\label{fig:tails_training}
\end{figure}

\begin{figure}\centering
\includegraphics[width=0.8\linewidth]{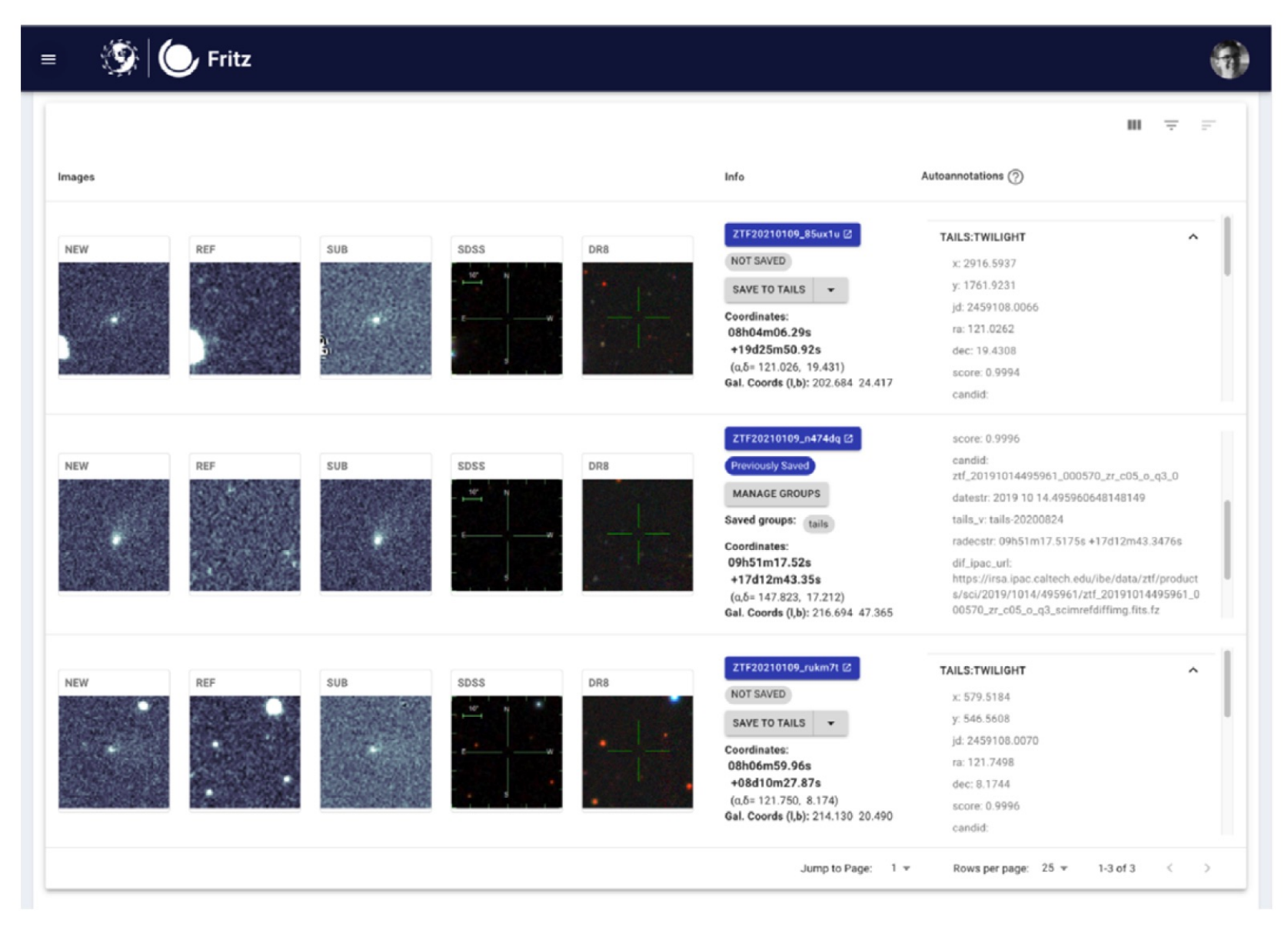} 
\caption{Example Tails interface for the scanning of comet candidates. Adapted from Fig.~5 of \citep[][]{Duev2021} and reproduced with the permission of the authors and AAS Journals.}
\label{fig:fritz}
\end{figure}

The performance of the Tails network with false positive and false negative rates as a function of the comet probability score is shown in Fig.~\ref{fig:tails_training}; both false positive and false negative rates are below 2\%. Tails has been run on the ZTF twilight survey data taken since 2019 September \citep[][]{Bolin2022IVO, Bolin2023Com}. The ZTF twilight survey consists of $\sim$50 images taken during both evening and morning nautical to astronomical twilight on a nightly basis with Sun-centric distances of 30-60 degrees \citep[][]{Bolin2022IVO,Bolin2024TS}. These near-Sun observations enable the detection of asteroids located inside the orbit of the Earth and Venus as well as of comets \citep[][]{Bolin2024TS}. It takes several hours of wall time on a 32-core machine to process a nightly data set. The output with comet probability scores, science, reference, and difference images are output to the Fritz user interface (see Fig.~\ref{fig:fritz}) \citep[][]{Coughlin2023Fritz}. 

\begin{figure}\centering
\includegraphics[width=0.8\linewidth]{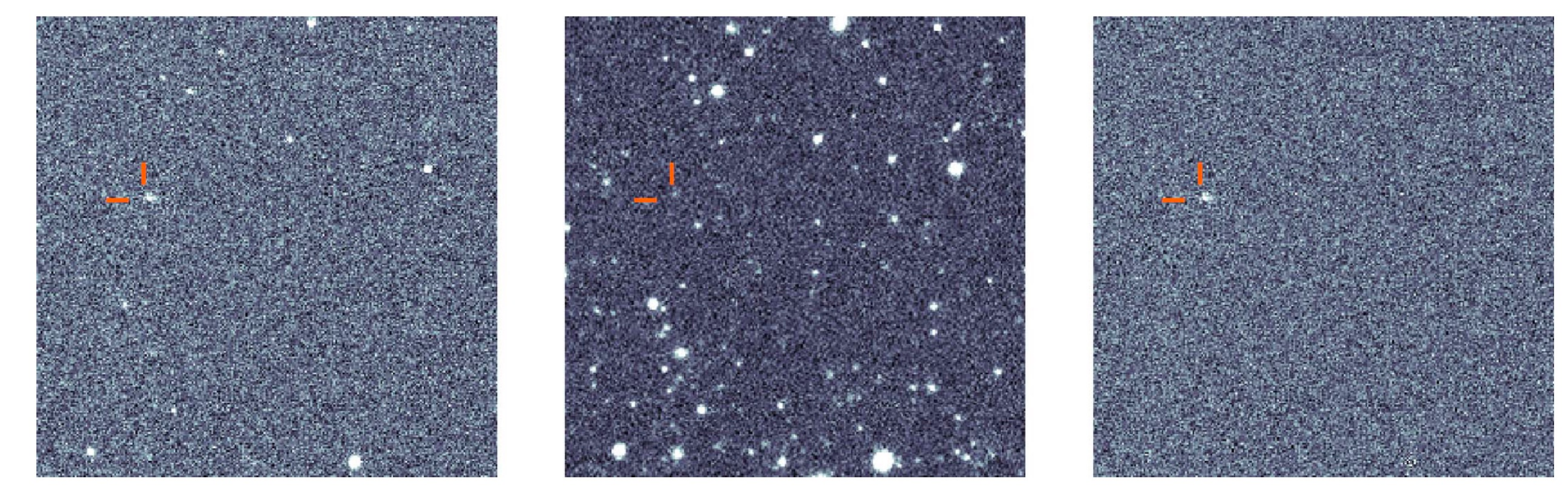} 
\caption{Discovery images of C/2020 T2 (Palomar) taken by ZTF on 2020 October 7. The left panel shows the science image, the center panel shows the reference image and the right panel shows the difference images  Adapted from Fig.~7 of \citep[][]{Duev2021} and reproduced with the permission of the authors and AAS Journals.}
\label{fig:T2}
\end{figure}

The first discovery of a comet made by AI-assisted methods through Tails was made in 2020 October of comet C/2020 T2 (Palomar) as seen in Fig.~\ref{fig:T2} \citep[][]{Duev2020MPECT2}. C/2020 T2 exhibits clear cometary features such as a tail and a coma $\sim$3-4 arcseconds wide compared to the seeing of $\sim$2 arcseconds. In addition to T2, Tails contributed to several other comet discoveries from the ZTF twilight survey, including C/2022 E3 \citep[][]{Bolin2022MPECE3} which will be described in a forthcoming publication \citep[][]{Bolin2022IVO,Bolin2024TS}. 

\subsection{Other applications of machine learning used to find comets and future developments}

Machine learning methods are becoming increasingly used in the detection and discovery of comets that have come out after the development of CNN-based methods like Tails \citep[e.g.,][]{Rozek2023, Sedaghat2024}. Future developments and improvements to CCN-based methods like Tails include the use of multiple-input CNNs, such as the BTSBot, implemented for identifying supernovae in ZTF images \citep[][]{Rehemtulla2023}. Multiple input methods combine additional features that can be used to enhance the identification of comets, such as orbital information and PSF-based information described in previous sections. By combining both CNN and other feature information, the accuracy of comet identification can be increased, which will be necessary for future surveys such as LSST.

\section*{Acknowledgments}

The authors wish to thank Dr. Valerio Carruba and Dr. Wesley Fraser for providing helpful reviews of this manuscript. Their comments were insightful in the improvement of the manuscript. The authors also wish to thank Ms. Laura-May Abron for her inspiration and artistic insight in the portrayal of ML-methods to identify solar system objects.



\bibliographystyle{elsarticle-harv} 
\bibliography{/Users/bolin/Dropbox/Projects/NEOZTF_NEOs/neobib}

\begin{thebibliography}{59}
\expandafter\ifx\csname natexlab\endcsname\relax\def\natexlab#1{#1}\fi
\providecommand{\url}[1]{\texttt{#1}}
\providecommand{\href}[2]{#2}
\providecommand{\path}[1]{#1}
\providecommand{\DOIprefix}{doi:}
\providecommand{\ArXivprefix}{arXiv:}
\providecommand{\URLprefix}{URL: }
\providecommand{\Pubmedprefix}{pmid:}
\providecommand{\doi}[1]{\href{http://dx.doi.org/#1}{\path{#1}}}
\providecommand{\Pubmed}[1]{\href{pmid:#1}{\path{#1}}}
\providecommand{\bibinfo}[2]{#2}
\ifx\xfnm\relax \def\xfnm[#1]{\unskip,\space#1}\fi
\bibitem[{{Battams} and {Knight}(2017)}]{Battams2017}
\bibinfo{author}{{Battams}, K.}, \bibinfo{author}{{Knight}, M.M.},
  \bibinfo{year}{2017}.
\newblock \bibinfo{title}{{SOHO comets: 20 years and 3000 objects later}}.
\newblock \bibinfo{journal}{Philosophical Transactions of the Royal Society of
  London Series A} \bibinfo{volume}{375}, \bibinfo{pages}{20160257}.
\newblock \DOIprefix\doi{10.1098/rsta.2016.0257},
  \href{http://arxiv.org/abs/1611.02279}{{\tt arXiv:1611.02279}}.
\bibitem[{{Bauer} et~al.(2022){Bauer}, {Fern{\'a}ndez}, {Protopapa} and
  {Woodney}}]{Bauer2022cometsurveys}
\bibinfo{author}{{Bauer}, J.M.}, \bibinfo{author}{{Fern{\'a}ndez}, Y.R.},
  \bibinfo{author}{{Protopapa}, S.}, \bibinfo{author}{{Woodney}, L.M.},
  \bibinfo{year}{2022}.
\newblock \bibinfo{title}{{Comet Science With Ground Based and Space Based
  Surveys in the New Millennium}}.
\newblock \bibinfo{journal}{arXiv e-prints} ,
  \bibinfo{pages}{arXiv:2210.09400}\DOIprefix\doi{10.48550/arXiv.2210.09400},
  \href{http://arxiv.org/abs/2210.09400}{{\tt arXiv:2210.09400}}.
\bibitem[{{Bellm} et~al.(2019){Bellm}, {Kulkarni}, {Graham}, {Dekany}, {Smith},
  {Riddle}, {Masci}, {Helou}, {Prince}, {Adams}, {Barbarino}, {Barlow},
  {Bauer}, {Beck}, {Belicki}, {Biswas}, {Blagorodnova}, {Bodewits}, {Bolin},
  {Brinnel}, {Brooke}, {Bue}, {Bulla}, {Burruss}, {Cenko}, {Chang}, {Connolly},
  {Coughlin}, {Cromer}, {Cunningham}, {De}, {Delacroix}, {Desai}, {Duev},
  {Eadie}, {Farnham}, {Feeney}, {Feindt}, {Flynn}, {Franckowiak}, {Frederick},
  {Fremling}, {Gal-Yam}, {Gezari}, {Giomi}, {Goldstein}, {Golkhou}, {Goobar},
  {Groom}, {Hacopians}, {Hale}, {Henning}, {Ho}, {Hover}, {Howell}, {Hung},
  {Huppenkothen}, {Imel}, {Ip}, {Ivezi{\'c}}, {Jackson}, {Jones}, {Juric},
  {Kasliwal}, {Kaspi}, {Kaye}, {Kelley}, {Kowalski}, {Kramer}, {Kupfer},
  {Landry}, {Laher}, {Lee}, {Lin}, {Lin}, {Lunnan}, {Giomi}, {Mahabal}, {Mao},
  {Miller}, {Monkewitz}, {Murphy}, {Ngeow}, {Nordin}, {Nugent}, {Ofek},
  {Patterson}, {Penprase}, {Porter}, {Rauch}, {Rebbapragada}, {Reiley},
  {Rigault}, {Rodriguez}, {van Roestel}, {Rusholme}, {van Santen}, {Schulze},
  {Shupe}, {Singer}, {Soumagnac}, {Stein}, {Surace}, {Sollerman}, {Szkody},
  {Taddia}, {Terek}, {Van Sistine}, {van Velzen}, {Vestrand}, {Walters},
  {Ward}, {Ye}, {Yu}, {Yan} and {Zolkower}}]{Bellm2019}
\bibinfo{author}{{Bellm}, E.C.}, \bibinfo{author}{{Kulkarni}, S.R.},
  \bibinfo{author}{{Graham}, M.J.}, \bibinfo{author}{{Dekany}, R.},
  \bibinfo{author}{{Smith}, R.M.}, \bibinfo{author}{{Riddle}, R.},
  \bibinfo{author}{{Masci}, F.J.}, \bibinfo{author}{{Helou}, G.},
  \bibinfo{author}{{Prince}, T.A.}, \bibinfo{author}{{Adams}, S.M.},
  \bibinfo{author}{{Barbarino}, C.}, \bibinfo{author}{{Barlow}, T.},
  \bibinfo{author}{{Bauer}, J.}, \bibinfo{author}{{Beck}, R.},
  \bibinfo{author}{{Belicki}, J.}, \bibinfo{author}{{Biswas}, R.},
  \bibinfo{author}{{Blagorodnova}, N.}, \bibinfo{author}{{Bodewits}, D.},
  \bibinfo{author}{{Bolin}, B.}, \bibinfo{author}{{Brinnel}, V.},
  \bibinfo{author}{{Brooke}, T.}, \bibinfo{author}{{Bue}, B.},
  \bibinfo{author}{{Bulla}, M.}, \bibinfo{author}{{Burruss}, R.},
  \bibinfo{author}{{Cenko}, S.B.}, \bibinfo{author}{{Chang}, C.K.},
  \bibinfo{author}{{Connolly}, A.}, \bibinfo{author}{{Coughlin}, M.},
  \bibinfo{author}{{Cromer}, J.}, \bibinfo{author}{{Cunningham}, V.},
  \bibinfo{author}{{De}, K.}, \bibinfo{author}{{Delacroix}, A.},
  \bibinfo{author}{{Desai}, V.}, \bibinfo{author}{{Duev}, D.A.},
  \bibinfo{author}{{Eadie}, G.}, \bibinfo{author}{{Farnham}, T.L.},
  \bibinfo{author}{{Feeney}, M.}, \bibinfo{author}{{Feindt}, U.},
  \bibinfo{author}{{Flynn}, D.}, \bibinfo{author}{{Franckowiak}, A.},
  \bibinfo{author}{{Frederick}, S.}, \bibinfo{author}{{Fremling}, C.},
  \bibinfo{author}{{Gal-Yam}, A.}, \bibinfo{author}{{Gezari}, S.},
  \bibinfo{author}{{Giomi}, M.}, \bibinfo{author}{{Goldstein}, D.A.},
  \bibinfo{author}{{Golkhou}, V.Z.}, \bibinfo{author}{{Goobar}, A.},
  \bibinfo{author}{{Groom}, S.}, \bibinfo{author}{{Hacopians}, E.},
  \bibinfo{author}{{Hale}, D.}, \bibinfo{author}{{Henning}, J.},
  \bibinfo{author}{{Ho}, A.Y.Q.}, \bibinfo{author}{{Hover}, D.},
  \bibinfo{author}{{Howell}, J.}, \bibinfo{author}{{Hung}, T.},
  \bibinfo{author}{{Huppenkothen}, D.}, \bibinfo{author}{{Imel}, D.},
  \bibinfo{author}{{Ip}, W.H.}, \bibinfo{author}{{Ivezi{\'c}}, {\v{Z}}.},
  \bibinfo{author}{{Jackson}, E.}, \bibinfo{author}{{Jones}, L.},
  \bibinfo{author}{{Juric}, M.}, \bibinfo{author}{{Kasliwal}, M.M.},
  \bibinfo{author}{{Kaspi}, S.}, \bibinfo{author}{{Kaye}, S.},
  \bibinfo{author}{{Kelley}, M.S.P.}, \bibinfo{author}{{Kowalski}, M.},
  \bibinfo{author}{{Kramer}, E.}, \bibinfo{author}{{Kupfer}, T.},
  \bibinfo{author}{{Landry}, W.}, \bibinfo{author}{{Laher}, R.R.},
  \bibinfo{author}{{Lee}, C.D.}, \bibinfo{author}{{Lin}, H.W.},
  \bibinfo{author}{{Lin}, Z.Y.}, \bibinfo{author}{{Lunnan}, R.},
  \bibinfo{author}{{Giomi}, M.}, \bibinfo{author}{{Mahabal}, A.},
  \bibinfo{author}{{Mao}, P.}, \bibinfo{author}{{Miller}, A.A.},
  \bibinfo{author}{{Monkewitz}, S.}, \bibinfo{author}{{Murphy}, P.},
  \bibinfo{author}{{Ngeow}, C.C.}, \bibinfo{author}{{Nordin}, J.},
  \bibinfo{author}{{Nugent}, P.}, \bibinfo{author}{{Ofek}, E.},
  \bibinfo{author}{{Patterson}, M.T.}, \bibinfo{author}{{Penprase}, B.},
  \bibinfo{author}{{Porter}, M.}, \bibinfo{author}{{Rauch}, L.},
  \bibinfo{author}{{Rebbapragada}, U.}, \bibinfo{author}{{Reiley}, D.},
  \bibinfo{author}{{Rigault}, M.}, \bibinfo{author}{{Rodriguez}, H.},
  \bibinfo{author}{{van Roestel}, J.}, \bibinfo{author}{{Rusholme}, B.},
  \bibinfo{author}{{van Santen}, J.}, \bibinfo{author}{{Schulze}, S.},
  \bibinfo{author}{{Shupe}, D.L.}, \bibinfo{author}{{Singer}, L.P.},
  \bibinfo{author}{{Soumagnac}, M.T.}, \bibinfo{author}{{Stein}, R.},
  \bibinfo{author}{{Surace}, J.}, \bibinfo{author}{{Sollerman}, J.},
  \bibinfo{author}{{Szkody}, P.}, \bibinfo{author}{{Taddia}, F.},
  \bibinfo{author}{{Terek}, S.}, \bibinfo{author}{{Van Sistine}, A.},
  \bibinfo{author}{{van Velzen}, S.}, \bibinfo{author}{{Vestrand}, W.T.},
  \bibinfo{author}{{Walters}, R.}, \bibinfo{author}{{Ward}, C.},
  \bibinfo{author}{{Ye}, Q.Z.}, \bibinfo{author}{{Yu}, P.C.},
  \bibinfo{author}{{Yan}, L.}, \bibinfo{author}{{Zolkower}, J.},
  \bibinfo{year}{2019}.
\newblock \bibinfo{title}{{The Zwicky Transient Facility: System Overview,
  Performance, and First Results}}.
\newblock \bibinfo{journal}{\pasp} \bibinfo{volume}{131},
  \bibinfo{pages}{018002}.
\newblock \DOIprefix\doi{10.1088/1538-3873/aaecbe}.
\bibitem[{{Bengio}(2009)}]{Bengio2009}
\bibinfo{author}{{Bengio}, Y.}, \bibinfo{year}{2009}.
\newblock \bibinfo{title}{{Learning Deep Architectures for AI}}.
\newblock \bibinfo{journal}{Foundation and Trends in AI} \bibinfo{volume}{2},
  \bibinfo{pages}{56}.
\bibitem[{{Bolin} et~al.(2013){Bolin}, {Denneau}, {Micheli}, {Wainscoat},
  {Tholen}, {Lister} and {Williams}}]{Bolin2013}
\bibinfo{author}{{Bolin}, B.}, \bibinfo{author}{{Denneau}, L.},
  \bibinfo{author}{{Micheli}, M.}, \bibinfo{author}{{Wainscoat}, R.},
  \bibinfo{author}{{Tholen}, D.J.}, \bibinfo{author}{{Lister}, T.},
  \bibinfo{author}{{Williams}, G.V.}, \bibinfo{year}{2013}.
\newblock \bibinfo{title}{{Comet P/2013 P5 (Panstarrs)}}.
\newblock \bibinfo{journal}{Central Bureau Electronic Telegrams}
  \bibinfo{volume}{3639}.
\bibitem[{{Bolin} et~al.(2023){Bolin}, {Ahumada}, {Dokkum}, {Fremling},
  {Hardegree-Ullman}, {Purdum}, {Serabyn} and {Southworth}}]{Bolin2023Com}
\bibinfo{author}{{Bolin}, B.T.}, \bibinfo{author}{{Ahumada}, T.},
  \bibinfo{author}{{Dokkum}, P.v.}, \bibinfo{author}{{Fremling}, C.},
  \bibinfo{author}{{Hardegree-Ullman}, K.K.}, \bibinfo{author}{{Purdum}, J.N.},
  \bibinfo{author}{{Serabyn}, E.}, \bibinfo{author}{{Southworth}, J.},
  \bibinfo{year}{2023}.
\newblock \bibinfo{title}{{Preliminary estimates of the Zwicky Transient
  Facility 'Ayl{\'o}'chaxnim asteroid population completeness}}.
\newblock \bibinfo{journal}{\icarus} \bibinfo{volume}{394},
  \bibinfo{pages}{115442}.
\newblock \DOIprefix\doi{10.1016/j.icarus.2023.115442}.
\bibitem[{{Bolin} et~al.(2022a){Bolin}, {Ahumada}, {van Dokkum}, {Fremling},
  {Granvik}, {Hardegree-Ullman}, {Harikane}, {Purdum}, {Serabyn}, {Southworth}
  and {Zhai}}]{Bolin2022IVO}
\bibinfo{author}{{Bolin}, B.T.}, \bibinfo{author}{{Ahumada}, T.},
  \bibinfo{author}{{van Dokkum}, P.}, \bibinfo{author}{{Fremling}, C.},
  \bibinfo{author}{{Granvik}, M.}, \bibinfo{author}{{Hardegree-Ullman}, K.K.},
  \bibinfo{author}{{Harikane}, Y.}, \bibinfo{author}{{Purdum}, J.N.},
  \bibinfo{author}{{Serabyn}, E.}, \bibinfo{author}{{Southworth}, J.},
  \bibinfo{author}{{Zhai}, C.}, \bibinfo{year}{2022}a.
\newblock \bibinfo{title}{{The discovery and characterization of (594913)
  'Ayl{\'o}'chaxnim, a kilometre sized asteroid inside the orbit of Venus}}.
\newblock \bibinfo{journal}{\mnras} \bibinfo{volume}{517},
  \bibinfo{pages}{L49--L54}.
\newblock \DOIprefix\doi{10.1093/mnrasl/slac089},
  \href{http://arxiv.org/abs/2208.07253}{{\tt arXiv:2208.07253}}.
\bibitem[{{Bolin} et~al.(2017){Bolin}, {Delbo}, {Morbidelli} and
  {Walsh}}]{Bolin2017}
\bibinfo{author}{{Bolin}, B.T.}, \bibinfo{author}{{Delbo}, M.},
  \bibinfo{author}{{Morbidelli}, A.}, \bibinfo{author}{{Walsh}, K.J.},
  \bibinfo{year}{2017}.
\newblock \bibinfo{title}{{Yarkovsky V-shape identification of asteroid
  families}}.
\newblock \bibinfo{journal}{\icarus} \bibinfo{volume}{282},
  \bibinfo{pages}{290--312}.
\newblock \DOIprefix\doi{10.1016/j.icarus.2016.09.029},
  \href{http://arxiv.org/abs/1609.06384}{{\tt arXiv:1609.06384}}.
\bibitem[{{Bolin} et~al.(2021){Bolin}, {Fernandez}, {Lisse}, {Holt}, {Lin},
  {Purdum}, {Deshmukh}, {Bauer}, {Bellm}, {Bodewits}, {Burdge}, {Carey},
  {Copperwheat}, {Helou}, {Ho}, {Horner}, {van Roestel}, {Bhalerao}, {Chang},
  {Chen}, {Hsu}, {Ip}, {Kasliwal}, {Masci}, {Ngeow}, {Quimby}, {Burruss},
  {Coughlin}, {Dekany}, {Delacroix}, {Drake}, {Duev}, {Graham}, {Hale},
  {Kupfer}, {Laher}, {Mahabal}, {Mr{\'o}z}, {Neill}, {Riddle}, {Rodriguez},
  {Smith}, {Soumagnac}, {Walters}, {Yan} and {Zolkower}}]{Bolin2021LD2}
\bibinfo{author}{{Bolin}, B.T.}, \bibinfo{author}{{Fernandez}, Y.R.},
  \bibinfo{author}{{Lisse}, C.M.}, \bibinfo{author}{{Holt}, T.R.},
  \bibinfo{author}{{Lin}, Z.Y.}, \bibinfo{author}{{Purdum}, J.N.},
  \bibinfo{author}{{Deshmukh}, K.P.}, \bibinfo{author}{{Bauer}, J.M.},
  \bibinfo{author}{{Bellm}, E.C.}, \bibinfo{author}{{Bodewits}, D.},
  \bibinfo{author}{{Burdge}, K.B.}, \bibinfo{author}{{Carey}, S.J.},
  \bibinfo{author}{{Copperwheat}, C.M.}, \bibinfo{author}{{Helou}, G.},
  \bibinfo{author}{{Ho}, A.Y.Q.}, \bibinfo{author}{{Horner}, J.},
  \bibinfo{author}{{van Roestel}, J.}, \bibinfo{author}{{Bhalerao}, V.},
  \bibinfo{author}{{Chang}, C.K.}, \bibinfo{author}{{Chen}, C.},
  \bibinfo{author}{{Hsu}, C.Y.}, \bibinfo{author}{{Ip}, W.H.},
  \bibinfo{author}{{Kasliwal}, M.M.}, \bibinfo{author}{{Masci}, F.J.},
  \bibinfo{author}{{Ngeow}, C.C.}, \bibinfo{author}{{Quimby}, R.},
  \bibinfo{author}{{Burruss}, R.}, \bibinfo{author}{{Coughlin}, M.},
  \bibinfo{author}{{Dekany}, R.}, \bibinfo{author}{{Delacroix}, A.},
  \bibinfo{author}{{Drake}, A.}, \bibinfo{author}{{Duev}, D.A.},
  \bibinfo{author}{{Graham}, M.}, \bibinfo{author}{{Hale}, D.},
  \bibinfo{author}{{Kupfer}, T.}, \bibinfo{author}{{Laher}, R.R.},
  \bibinfo{author}{{Mahabal}, A.}, \bibinfo{author}{{Mr{\'o}z}, P.J.},
  \bibinfo{author}{{Neill}, J.D.}, \bibinfo{author}{{Riddle}, R.},
  \bibinfo{author}{{Rodriguez}, H.}, \bibinfo{author}{{Smith}, R.M.},
  \bibinfo{author}{{Soumagnac}, M.T.}, \bibinfo{author}{{Walters}, R.},
  \bibinfo{author}{{Yan}, L.}, \bibinfo{author}{{Zolkower}, J.},
  \bibinfo{year}{2021}.
\newblock \bibinfo{title}{{Initial Characterization of Active Transitioning
  Centaur, P/2019 LD$_{2}$ (ATLAS), Using Hubble, Spitzer, ZTF, Keck, Apache
  Point Observatory, and GROWTH Visible and Infrared Imaging and
  Spectroscopy}}.
\newblock \bibinfo{journal}{\aj} \bibinfo{volume}{161}, \bibinfo{pages}{116}.
\newblock \DOIprefix\doi{10.3847/1538-3881/abd94b},
  \href{http://arxiv.org/abs/2011.03782}{{\tt arXiv:2011.03782}}.
\bibitem[{{Bolin} et~al.(2020){Bolin}, {Fremling}, {Holt}, {Hankins},
  {Ahumada}, {Anand}, {Bhalerao}, {Burdge}, {Copperwheat}, {Coughlin},
  {Deshmukh}, {De}, {Kasliwal}, {Morbidelli}, {Purdum}, {Quimby}, {Bodewits},
  {Chang}, {Ip}, {Hsu}, {Laher}, {Lin}, {Lisse}, {Masci}, {Ngeow}, {Tan},
  {Zhai}, {Burruss}, {Dekany}, {Delacroix}, {Duev}, {Graham}, {Hale},
  {Kulkarni}, {Kupfer}, {Mahabal}, {Mr{\'o}z}, {Neill}, {Riddle}, {Rodriguez},
  {Smith}, {Soumagnac}, {Walters}, {Yan} and {Zolkower}}]{Bolin2020CD3}
\bibinfo{author}{{Bolin}, B.T.}, \bibinfo{author}{{Fremling}, C.},
  \bibinfo{author}{{Holt}, T.R.}, \bibinfo{author}{{Hankins}, M.J.},
  \bibinfo{author}{{Ahumada}, T.}, \bibinfo{author}{{Anand}, S.},
  \bibinfo{author}{{Bhalerao}, V.}, \bibinfo{author}{{Burdge}, K.B.},
  \bibinfo{author}{{Copperwheat}, C.M.}, \bibinfo{author}{{Coughlin}, M.},
  \bibinfo{author}{{Deshmukh}, K.P.}, \bibinfo{author}{{De}, K.},
  \bibinfo{author}{{Kasliwal}, M.M.}, \bibinfo{author}{{Morbidelli}, A.},
  \bibinfo{author}{{Purdum}, J.N.}, \bibinfo{author}{{Quimby}, R.},
  \bibinfo{author}{{Bodewits}, D.}, \bibinfo{author}{{Chang}, C.K.},
  \bibinfo{author}{{Ip}, W.H.}, \bibinfo{author}{{Hsu}, C.Y.},
  \bibinfo{author}{{Laher}, R.R.}, \bibinfo{author}{{Lin}, Z.Y.},
  \bibinfo{author}{{Lisse}, C.M.}, \bibinfo{author}{{Masci}, F.J.},
  \bibinfo{author}{{Ngeow}, C.C.}, \bibinfo{author}{{Tan}, H.},
  \bibinfo{author}{{Zhai}, C.}, \bibinfo{author}{{Burruss}, R.},
  \bibinfo{author}{{Dekany}, R.}, \bibinfo{author}{{Delacroix}, A.},
  \bibinfo{author}{{Duev}, D.A.}, \bibinfo{author}{{Graham}, M.},
  \bibinfo{author}{{Hale}, D.}, \bibinfo{author}{{Kulkarni}, S.R.},
  \bibinfo{author}{{Kupfer}, T.}, \bibinfo{author}{{Mahabal}, A.},
  \bibinfo{author}{{Mr{\'o}z}, P.J.}, \bibinfo{author}{{Neill}, J.D.},
  \bibinfo{author}{{Riddle}, R.}, \bibinfo{author}{{Rodriguez}, H.},
  \bibinfo{author}{{Smith}, R.M.}, \bibinfo{author}{{Soumagnac}, M.T.},
  \bibinfo{author}{{Walters}, R.}, \bibinfo{author}{{Yan}, L.},
  \bibinfo{author}{{Zolkower}, J.}, \bibinfo{year}{2020}.
\newblock \bibinfo{title}{{Characterization of Temporarily Captured Minimoon
  2020 CD$_{3}$ by Keck Time-resolved Spectrophotometry}}.
\newblock \bibinfo{journal}{\apjl} \bibinfo{volume}{900}, \bibinfo{pages}{L45}.
\newblock \DOIprefix\doi{10.3847/2041-8213/abae69},
  \href{http://arxiv.org/abs/2008.05384}{{\tt arXiv:2008.05384}}.
\bibitem[{{Bolin} and {Lisse}(2020)}]{Bolin2020HST}
\bibinfo{author}{{Bolin}, B.T.}, \bibinfo{author}{{Lisse}, C.M.},
  \bibinfo{year}{2020}.
\newblock \bibinfo{title}{{Constraints on the spin-pole orientation, jet
  morphology, and rotation of interstellar comet 2I/Borisov with deep HST
  imaging}}.
\newblock \bibinfo{journal}{\mnras} \bibinfo{volume}{497},
  \bibinfo{pages}{4031--4041}.
\newblock \DOIprefix\doi{10.1093/mnras/staa2192},
  \href{http://arxiv.org/abs/1912.07386}{{\tt arXiv:1912.07386}}.
\bibitem[{{Bolin} et~al.(2024a){Bolin}, {Masci}, {Coughlin}, {Duev} and
  Stubbs}]{Bolin2024TS}
\bibinfo{author}{{Bolin}, B.T.}, \bibinfo{author}{{Masci}, F.J.},
  \bibinfo{author}{{Coughlin}, M.W.}, \bibinfo{author}{{Duev}, D.W.},
  \bibinfo{author}{Stubbs, C.W.}, \bibinfo{year}{2024}a.
\newblock \bibinfo{title}{{An artificial-intelligence powered twilight survey
  of 'Ayl\'{o}'chaxnim, Atiras, and comets at Palomar Observatory}}.
\newblock \bibinfo{journal}{Icarus, submitted.} .
\bibitem[{{Bolin} et~al.(2024b){Bolin}, {Masci}, {Duev}, {Milburn}, {Neill},
  {Purdum}, {Avdellidou}, {Saki}, {Cheng}, {Delbo}, {Fremling}, {Ghosal},
  {Lin}, {Lisse} and {Mahabal}}]{Bolin2024E3}
\bibinfo{author}{{Bolin}, B.T.}, \bibinfo{author}{{Masci}, F.J.},
  \bibinfo{author}{{Duev}, D.A.}, \bibinfo{author}{{Milburn}, J.W.},
  \bibinfo{author}{{Neill}, J.D.}, \bibinfo{author}{{Purdum}, J.N.},
  \bibinfo{author}{{Avdellidou}, C.}, \bibinfo{author}{{Saki}, M.},
  \bibinfo{author}{{Cheng}, Y.C.}, \bibinfo{author}{{Delbo}, M.},
  \bibinfo{author}{{Fremling}, C.}, \bibinfo{author}{{Ghosal}, M.},
  \bibinfo{author}{{Lin}, Z.Y.}, \bibinfo{author}{{Lisse}, C.M.},
  \bibinfo{author}{{Mahabal}, A.}, \bibinfo{year}{2024}b.
\newblock \bibinfo{title}{{Palomar discovery and initial characterization of
  naked-eye long-period comet C/2022 E3 (ZTF)}}.
\newblock \bibinfo{journal}{\mnras} \bibinfo{volume}{527},
  \bibinfo{pages}{L42--L46}.
\newblock \DOIprefix\doi{10.1093/mnrasl/slad139},
  \href{http://arxiv.org/abs/2309.14336}{{\tt arXiv:2309.14336}}.
\bibitem[{{Bolin} et~al.(2022b){Bolin}, {Masci}, {Ip}, {Helou}, {Kramer},
  {Lin}, {Prince}, {Sato}, {Paul}, {Yoshimoto}, {Urbanik}, {Denneau}, {Siverd},
  {Tonry}, {Weiland}, {Erasmus}, {Fitzsimmons}, {Lawrence}, {Robinson},
  {Siverd}, {Tonry}, {Birtwhistle}, {Jacques}, {Hug}, {Korlevic}, {Buzzi},
  {Bacci}, {van Buitenen}, {Buczynski}, {Hale}, {Masek}, {Guido}, {Rocchetto},
  {Bryssinck}, {Milani}, {Savini}, {Valvasori}, {Ligustri}, {Bacci},
  {Maestripieri}, {Tesi}, {Fagioli} and {Lutkenhoner}}]{Bolin2022MPECE3}
\bibinfo{author}{{Bolin}, B.T.}, \bibinfo{author}{{Masci}, F.J.},
  \bibinfo{author}{{Ip}, W.H.}, \bibinfo{author}{{Helou}, G.},
  \bibinfo{author}{{Kramer}, E.A.}, \bibinfo{author}{{Lin}, Z.Y.},
  \bibinfo{author}{{Prince}, T.A.}, \bibinfo{author}{{Sato}, H.},
  \bibinfo{author}{{Paul}, N.}, \bibinfo{author}{{Yoshimoto}, K.},
  \bibinfo{author}{{Urbanik}, M.}, \bibinfo{author}{{Denneau}, L.},
  \bibinfo{author}{{Siverd}, R.}, \bibinfo{author}{{Tonry}, J.},
  \bibinfo{author}{{Weiland}, H.}, \bibinfo{author}{{Erasmus}, N.},
  \bibinfo{author}{{Fitzsimmons}, A.}, \bibinfo{author}{{Lawrence}, A.},
  \bibinfo{author}{{Robinson}, J.}, \bibinfo{author}{{Siverd}, R.},
  \bibinfo{author}{{Tonry}, J.}, \bibinfo{author}{{Birtwhistle}, P.},
  \bibinfo{author}{{Jacques}, C.}, \bibinfo{author}{{Hug}, G.},
  \bibinfo{author}{{Korlevic}, K.}, \bibinfo{author}{{Buzzi}, L.},
  \bibinfo{author}{{Bacci}, R.}, \bibinfo{author}{{van Buitenen}, G.},
  \bibinfo{author}{{Buczynski}, D.}, \bibinfo{author}{{Hale}, A.},
  \bibinfo{author}{{Masek}, M.}, \bibinfo{author}{{Guido}, E.},
  \bibinfo{author}{{Rocchetto}, M.}, \bibinfo{author}{{Bryssinck}, E.},
  \bibinfo{author}{{Milani}, G.}, \bibinfo{author}{{Savini}, G.},
  \bibinfo{author}{{Valvasori}, A.}, \bibinfo{author}{{Ligustri}, R.},
  \bibinfo{author}{{Bacci}, P.}, \bibinfo{author}{{Maestripieri}, M.},
  \bibinfo{author}{{Tesi}, L.}, \bibinfo{author}{{Fagioli}, G.},
  \bibinfo{author}{{Lutkenhoner}, B.}, \bibinfo{year}{2022}b.
\newblock \bibinfo{title}{{Comet C/2022 E3 (ZTF)}}.
\newblock \bibinfo{journal}{Minor Planet Electronic Circulars}
  \bibinfo{volume}{2022-F13}.
\bibitem[{{Bolin} et~al.(2018){Bolin}, {Morbidelli} and {Walsh}}]{Bolin2017b}
\bibinfo{author}{{Bolin}, B.T.}, \bibinfo{author}{{Morbidelli}, A.},
  \bibinfo{author}{{Walsh}, K.J.}, \bibinfo{year}{2018}.
\newblock \bibinfo{title}{{Size-dependent modification of asteroid family
  Yarkovsky V-shapes}}.
\newblock \bibinfo{journal}{\aap} \bibinfo{volume}{611}, \bibinfo{pages}{A82}.
\newblock \DOIprefix\doi{10.1051/0004-6361/201732079},
  \href{http://arxiv.org/abs/1710.04208}{{\tt arXiv:1710.04208}}.
\bibitem[{{Born} and {Wolf}(1999)}]{Born1999}
\bibinfo{author}{{Born}, M.}, \bibinfo{author}{{Wolf}, E.},
  \bibinfo{year}{1999}.
\newblock \bibinfo{title}{{Principles of Optics}}.
\bibitem[{{Bowell} et~al.(1988){Bowell}, {Hapke}, {Domingue}, {Lumme},
  {Peltoniemi} and {Harris}}]{Bowell1988}
\bibinfo{author}{{Bowell}, E.}, \bibinfo{author}{{Hapke}, B.},
  \bibinfo{author}{{Domingue}, D.}, \bibinfo{author}{{Lumme}, K.},
  \bibinfo{author}{{Peltoniemi}, J.}, \bibinfo{author}{{Harris}, A.},
  \bibinfo{year}{1988}.
\newblock \bibinfo{title}{{Application of Photometric Models to Asteroids}}.
\newblock \bibinfo{journal}{Asteroids II} , \bibinfo{pages}{399--433}.
\bibitem[{{Chambers} et~al.(2016){Chambers}, {Magnier}, {Metcalfe},
  {Flewelling}, {Huber}, {Waters}, {Denneau}, {Draper}, {Farrow}, {Finkbeiner},
  {Holmberg}, {Koppenhoefer}, {Price}, {Saglia}, {Schlafly}, {Smartt},
  {Sweeney}, {Wainscoat}, {Burgett}, {Grav}, {Heasley}, {Hodapp}, {Jedicke},
  {Kaiser}, {Kudritzki}, {Luppino}, {Lupton}, {Monet}, {Morgan}, {Onaka},
  {Stubbs}, {Tonry}, {Banados}, {Bell}, {Bender}, {Bernard}, {Botticella},
  {Casertano}, {Chastel}, {Chen}, {Chen}, {Cole}, {Deacon}, {Frenk},
  {Fitzsimmons}, {Gezari}, {Goessl}, {Goggia}, {Goldman}, {Grebel}, {Hambly},
  {Hasinger}, {Heavens}, {Heckman}, {Henderson}, {Henning}, {Holman}, {Hopp},
  {Ip}, {Isani}, {Keyes}, {Koekemoer}, {Kotak}, {Long}, {Lucey}, {Liu},
  {Martin}, {McLean}, {Morganson}, {Murphy}, {Nieto-Santisteban}, {Norberg},
  {Peacock}, {Pier}, {Postman}, {Primak}, {Rae}, {Rest}, {Riess}, {Riffeser},
  {Rix}, {Roser}, {Schilbach}, {Schultz}, {Scolnic}, {Szalay}, {Seitz},
  {Shiao}, {Small}, {Smith}, {Soderblom}, {Taylor}, {Thakar}, {Thiel},
  {Thilker}, {Urata}, {Valenti}, {Walter}, {Watters}, {Werner}, {White},
  {Wood-Vasey} and {Wyse}}]{Chambers2016}
\bibinfo{author}{{Chambers}, K.C.}, \bibinfo{author}{{Magnier}, E.A.},
  \bibinfo{author}{{Metcalfe}, N.}, \bibinfo{author}{{Flewelling}, H.A.},
  \bibinfo{author}{{Huber}, M.E.}, \bibinfo{author}{{Waters}, C.Z.},
  \bibinfo{author}{{Denneau}, L.}, \bibinfo{author}{{Draper}, P.W.},
  \bibinfo{author}{{Farrow}, D.}, \bibinfo{author}{{Finkbeiner}, D.P.},
  \bibinfo{author}{{Holmberg}, C.}, \bibinfo{author}{{Koppenhoefer}, J.},
  \bibinfo{author}{{Price}, P.A.}, \bibinfo{author}{{Saglia}, R.P.},
  \bibinfo{author}{{Schlafly}, E.F.}, \bibinfo{author}{{Smartt}, S.J.},
  \bibinfo{author}{{Sweeney}, W.}, \bibinfo{author}{{Wainscoat}, R.J.},
  \bibinfo{author}{{Burgett}, W.S.}, \bibinfo{author}{{Grav}, T.},
  \bibinfo{author}{{Heasley}, J.N.}, \bibinfo{author}{{Hodapp}, K.W.},
  \bibinfo{author}{{Jedicke}, R.}, \bibinfo{author}{{Kaiser}, N.},
  \bibinfo{author}{{Kudritzki}, R.P.}, \bibinfo{author}{{Luppino}, G.A.},
  \bibinfo{author}{{Lupton}, R.H.}, \bibinfo{author}{{Monet}, D.G.},
  \bibinfo{author}{{Morgan}, J.S.}, \bibinfo{author}{{Onaka}, P.M.},
  \bibinfo{author}{{Stubbs}, C.W.}, \bibinfo{author}{{Tonry}, J.L.},
  \bibinfo{author}{{Banados}, E.}, \bibinfo{author}{{Bell}, E.F.},
  \bibinfo{author}{{Bender}, R.}, \bibinfo{author}{{Bernard}, E.J.},
  \bibinfo{author}{{Botticella}, M.T.}, \bibinfo{author}{{Casertano}, S.},
  \bibinfo{author}{{Chastel}, S.}, \bibinfo{author}{{Chen}, W.P.},
  \bibinfo{author}{{Chen}, X.}, \bibinfo{author}{{Cole}, S.},
  \bibinfo{author}{{Deacon}, N.}, \bibinfo{author}{{Frenk}, C.},
  \bibinfo{author}{{Fitzsimmons}, A.}, \bibinfo{author}{{Gezari}, S.},
  \bibinfo{author}{{Goessl}, C.}, \bibinfo{author}{{Goggia}, T.},
  \bibinfo{author}{{Goldman}, B.}, \bibinfo{author}{{Grebel}, E.K.},
  \bibinfo{author}{{Hambly}, N.C.}, \bibinfo{author}{{Hasinger}, G.},
  \bibinfo{author}{{Heavens}, A.F.}, \bibinfo{author}{{Heckman}, T.M.},
  \bibinfo{author}{{Henderson}, R.}, \bibinfo{author}{{Henning}, T.},
  \bibinfo{author}{{Holman}, M.}, \bibinfo{author}{{Hopp}, U.},
  \bibinfo{author}{{Ip}, W.H.}, \bibinfo{author}{{Isani}, S.},
  \bibinfo{author}{{Keyes}, C.D.}, \bibinfo{author}{{Koekemoer}, A.},
  \bibinfo{author}{{Kotak}, R.}, \bibinfo{author}{{Long}, K.S.},
  \bibinfo{author}{{Lucey}, J.R.}, \bibinfo{author}{{Liu}, M.},
  \bibinfo{author}{{Martin}, N.F.}, \bibinfo{author}{{McLean}, B.},
  \bibinfo{author}{{Morganson}, E.}, \bibinfo{author}{{Murphy}, D.N.A.},
  \bibinfo{author}{{Nieto-Santisteban}, M.A.}, \bibinfo{author}{{Norberg}, P.},
  \bibinfo{author}{{Peacock}, J.A.}, \bibinfo{author}{{Pier}, E.A.},
  \bibinfo{author}{{Postman}, M.}, \bibinfo{author}{{Primak}, N.},
  \bibinfo{author}{{Rae}, C.}, \bibinfo{author}{{Rest}, A.},
  \bibinfo{author}{{Riess}, A.}, \bibinfo{author}{{Riffeser}, A.},
  \bibinfo{author}{{Rix}, H.W.}, \bibinfo{author}{{Roser}, S.},
  \bibinfo{author}{{Schilbach}, E.}, \bibinfo{author}{{Schultz}, A.S.B.},
  \bibinfo{author}{{Scolnic}, D.}, \bibinfo{author}{{Szalay}, A.},
  \bibinfo{author}{{Seitz}, S.}, \bibinfo{author}{{Shiao}, B.},
  \bibinfo{author}{{Small}, E.}, \bibinfo{author}{{Smith}, K.W.},
  \bibinfo{author}{{Soderblom}, D.}, \bibinfo{author}{{Taylor}, A.N.},
  \bibinfo{author}{{Thakar}, A.R.}, \bibinfo{author}{{Thiel}, J.},
  \bibinfo{author}{{Thilker}, D.}, \bibinfo{author}{{Urata}, Y.},
  \bibinfo{author}{{Valenti}, J.}, \bibinfo{author}{{Walter}, F.},
  \bibinfo{author}{{Watters}, S.P.}, \bibinfo{author}{{Werner}, S.},
  \bibinfo{author}{{White}, R.}, \bibinfo{author}{{Wood-Vasey}, W.M.},
  \bibinfo{author}{{Wyse}, R.}, \bibinfo{year}{2016}.
\newblock \bibinfo{title}{{The Pan-STARRS1 Surveys}}.
\newblock \bibinfo{journal}{ArXiv e-prints}
  \href{http://arxiv.org/abs/1612.05560}{{\tt arXiv:1612.05560}}.
\bibitem[{{Chandler} et~al.(2021){Chandler}, {Trujillo} and
  {Hsieh}}]{Chandler2021}
\bibinfo{author}{{Chandler}, C.O.}, \bibinfo{author}{{Trujillo}, C.A.},
  \bibinfo{author}{{Hsieh}, H.H.}, \bibinfo{year}{2021}.
\newblock \bibinfo{title}{{Recurrent Activity from Active Asteroid (248370)
  2005 QN$_{173}$: A Main-belt Comet}}.
\newblock \bibinfo{journal}{\apjl} \bibinfo{volume}{922}, \bibinfo{pages}{L8}.
\newblock \DOIprefix\doi{10.3847/2041-8213/ac365b},
  \href{http://arxiv.org/abs/2111.06405}{{\tt arXiv:2111.06405}}.
\bibitem[{{Cheng} and {Wu}(2022)}]{Cheng2022A}
\bibinfo{author}{{Cheng}, Y.C.}, \bibinfo{author}{{Wu}, Y.L.},
  \bibinfo{year}{2022}.
\newblock \bibinfo{title}{{Cometary Activities of the Hyperbolic Asteroid
  A/2021 X2 Observed at Lulin Observatory}}.
\newblock \bibinfo{journal}{The Astronomer's Telegram} \bibinfo{volume}{15597},
  \bibinfo{pages}{1}.
\bibitem[{{Chyba Rabeendran} and {Denneau}(2021)}]{Rabeendran2021}
\bibinfo{author}{{Chyba Rabeendran}, A.}, \bibinfo{author}{{Denneau}, L.},
  \bibinfo{year}{2021}.
\newblock \bibinfo{title}{{A Two-stage Deep Learning Detection Classifier for
  the ATLAS Asteroid Survey}}.
\newblock \bibinfo{journal}{\pasp} \bibinfo{volume}{133},
  \bibinfo{pages}{034501}.
\newblock \DOIprefix\doi{10.1088/1538-3873/abc900},
  \href{http://arxiv.org/abs/2101.08912}{{\tt arXiv:2101.08912}}.
\bibitem[{{Coughlin} et~al.(2023){Coughlin}, {Bloom}, {Nir}, {Antier}, {du
  Laz}, {van der Walt}, {Crellin-Quick}, {Culino}, {Duev}, {Goldstein},
  {Healy}, {Karambelkar}, {Lilleboe}, {Shin}, {Singer}, {Ahumada}, {Anand},
  {Bellm}, {Dekany}, {Graham}, {Kasliwal}, {Kostadinova}, {Kiendrebeogo},
  {Kulkarni}, {Jenkins}, {LeBaron}, {Mahabal}, {Neill}, {Parazin}, {Peloton},
  {Perley}, {Riddle}, {Rusholme}, {van Santen}, {Sollerman}, {Stein}, {Turpin},
  {Wold}, {Amat}, {Bonnefon}, {Bonnefoy}, {Flament}, {Kerkow}, {Kishore},
  {Jani}, {Mahanty}, {Liu}, {Llinares}, {Makarison}, {Olli{\'e}ric}, {Perez},
  {Pont} and {Sharma}}]{Coughlin2023Fritz}
\bibinfo{author}{{Coughlin}, M.W.}, \bibinfo{author}{{Bloom}, J.S.},
  \bibinfo{author}{{Nir}, G.}, \bibinfo{author}{{Antier}, S.},
  \bibinfo{author}{{du Laz}, T.J.}, \bibinfo{author}{{van der Walt}, S.},
  \bibinfo{author}{{Crellin-Quick}, A.}, \bibinfo{author}{{Culino}, T.},
  \bibinfo{author}{{Duev}, D.A.}, \bibinfo{author}{{Goldstein}, D.A.},
  \bibinfo{author}{{Healy}, B.F.}, \bibinfo{author}{{Karambelkar}, V.},
  \bibinfo{author}{{Lilleboe}, J.}, \bibinfo{author}{{Shin}, K.M.},
  \bibinfo{author}{{Singer}, L.P.}, \bibinfo{author}{{Ahumada}, T.},
  \bibinfo{author}{{Anand}, S.}, \bibinfo{author}{{Bellm}, E.C.},
  \bibinfo{author}{{Dekany}, R.}, \bibinfo{author}{{Graham}, M.J.},
  \bibinfo{author}{{Kasliwal}, M.M.}, \bibinfo{author}{{Kostadinova}, I.},
  \bibinfo{author}{{Kiendrebeogo}, R.W.}, \bibinfo{author}{{Kulkarni}, S.R.},
  \bibinfo{author}{{Jenkins}, S.}, \bibinfo{author}{{LeBaron}, N.},
  \bibinfo{author}{{Mahabal}, A.A.}, \bibinfo{author}{{Neill}, J.D.},
  \bibinfo{author}{{Parazin}, B.}, \bibinfo{author}{{Peloton}, J.},
  \bibinfo{author}{{Perley}, D.A.}, \bibinfo{author}{{Riddle}, R.},
  \bibinfo{author}{{Rusholme}, B.}, \bibinfo{author}{{van Santen}, J.},
  \bibinfo{author}{{Sollerman}, J.}, \bibinfo{author}{{Stein}, R.},
  \bibinfo{author}{{Turpin}, D.}, \bibinfo{author}{{Wold}, A.},
  \bibinfo{author}{{Amat}, C.}, \bibinfo{author}{{Bonnefon}, A.},
  \bibinfo{author}{{Bonnefoy}, A.}, \bibinfo{author}{{Flament}, M.},
  \bibinfo{author}{{Kerkow}, F.}, \bibinfo{author}{{Kishore}, S.},
  \bibinfo{author}{{Jani}, S.}, \bibinfo{author}{{Mahanty}, S.K.},
  \bibinfo{author}{{Liu}, C.}, \bibinfo{author}{{Llinares}, L.},
  \bibinfo{author}{{Makarison}, J.}, \bibinfo{author}{{Olli{\'e}ric}, A.},
  \bibinfo{author}{{Perez}, I.}, \bibinfo{author}{{Pont}, L.},
  \bibinfo{author}{{Sharma}, V.}, \bibinfo{year}{2023}.
\newblock \bibinfo{title}{{A Data Science Platform to Enable Time-domain
  Astronomy}}.
\newblock \bibinfo{journal}{\apjs} \bibinfo{volume}{267}, \bibinfo{pages}{31}.
\newblock \DOIprefix\doi{10.3847/1538-4365/acdee1},
  \href{http://arxiv.org/abs/2305.00108}{{\tt arXiv:2305.00108}}.
\bibitem[{{Denneau} et~al.(2013){Denneau}, {Jedicke}, {Grav}, {Granvik},
  {Kubica}, {Milani}, {Vere{\v s}}, {Wainscoat}, {Chang}, {Pierfederici},
  {Kaiser}, {Chambers}, {Heasley}, {Magnier}, {Price}, {Myers}, {Kleyna},
  {Hsieh}, {Farnocchia}, {Waters}, {Sweeney}, {Green}, {Bolin}, {Burgett},
  {Morgan}, {Tonry}, {Hodapp}, {Chastel}, {Chesley}, {Fitzsimmons}, {Holman},
  {Spahr}, {Tholen}, {Williams}, {Abe}, {Armstrong}, {Bressi}, {Holmes},
  {Lister}, {McMillan}, {Micheli}, {Ryan}, {Ryan} and {Scotti}}]{Denneau2013}
\bibinfo{author}{{Denneau}, L.}, \bibinfo{author}{{Jedicke}, R.},
  \bibinfo{author}{{Grav}, T.}, \bibinfo{author}{{Granvik}, M.},
  \bibinfo{author}{{Kubica}, J.}, \bibinfo{author}{{Milani}, A.},
  \bibinfo{author}{{Vere{\v s}}, P.}, \bibinfo{author}{{Wainscoat}, R.},
  \bibinfo{author}{{Chang}, D.}, \bibinfo{author}{{Pierfederici}, F.},
  \bibinfo{author}{{Kaiser}, N.}, \bibinfo{author}{{Chambers}, K.C.},
  \bibinfo{author}{{Heasley}, J.N.}, \bibinfo{author}{{Magnier}, E.A.},
  \bibinfo{author}{{Price}, P.A.}, \bibinfo{author}{{Myers}, J.},
  \bibinfo{author}{{Kleyna}, J.}, \bibinfo{author}{{Hsieh}, H.},
  \bibinfo{author}{{Farnocchia}, D.}, \bibinfo{author}{{Waters}, C.},
  \bibinfo{author}{{Sweeney}, W.H.}, \bibinfo{author}{{Green}, D.},
  \bibinfo{author}{{Bolin}, B.}, \bibinfo{author}{{Burgett}, W.S.},
  \bibinfo{author}{{Morgan}, J.S.}, \bibinfo{author}{{Tonry}, J.L.},
  \bibinfo{author}{{Hodapp}, K.W.}, \bibinfo{author}{{Chastel}, S.},
  \bibinfo{author}{{Chesley}, S.}, \bibinfo{author}{{Fitzsimmons}, A.},
  \bibinfo{author}{{Holman}, M.}, \bibinfo{author}{{Spahr}, T.},
  \bibinfo{author}{{Tholen}, D.}, \bibinfo{author}{{Williams}, G.V.},
  \bibinfo{author}{{Abe}, S.}, \bibinfo{author}{{Armstrong}, J.D.},
  \bibinfo{author}{{Bressi}, T.H.}, \bibinfo{author}{{Holmes}, R.},
  \bibinfo{author}{{Lister}, T.}, \bibinfo{author}{{McMillan}, R.S.},
  \bibinfo{author}{{Micheli}, M.}, \bibinfo{author}{{Ryan}, E.V.},
  \bibinfo{author}{{Ryan}, W.H.}, \bibinfo{author}{{Scotti}, J.V.},
  \bibinfo{year}{2013}.
\newblock \bibinfo{title}{{The Pan-STARRS Moving Object Processing System}}.
\newblock \bibinfo{journal}{\pasp} \bibinfo{volume}{125},
  \bibinfo{pages}{357--395}.
\newblock \DOIprefix\doi{10.1086/670337},
  \href{http://arxiv.org/abs/1302.7281}{{\tt arXiv:1302.7281}}.
\bibitem[{{Dobson} et~al.(2023){Dobson}, {Schwamb}, {Benecchi}, {Verbiscer},
  {Fitzsimmons}, {Shingles}, {Denneau}, {Heinze}, {Smith}, {Tonry}, {Weiland}
  and {Young}}]{Dobson2023}
\bibinfo{author}{{Dobson}, M.M.}, \bibinfo{author}{{Schwamb}, M.E.},
  \bibinfo{author}{{Benecchi}, S.D.}, \bibinfo{author}{{Verbiscer}, A.J.},
  \bibinfo{author}{{Fitzsimmons}, A.}, \bibinfo{author}{{Shingles}, L.J.},
  \bibinfo{author}{{Denneau}, L.}, \bibinfo{author}{{Heinze}, A.N.},
  \bibinfo{author}{{Smith}, K.W.}, \bibinfo{author}{{Tonry}, J.L.},
  \bibinfo{author}{{Weiland}, H.}, \bibinfo{author}{{Young}, D.R.},
  \bibinfo{year}{2023}.
\newblock \bibinfo{title}{{Phase Curves of Kuiper Belt Objects, Centaurs, and
  Jupiter-family Comets from the ATLAS Survey}}.
\newblock \bibinfo{journal}{\psj} \bibinfo{volume}{4}, \bibinfo{pages}{75}.
\newblock \DOIprefix\doi{10.3847/PSJ/acc463},
  \href{http://arxiv.org/abs/2303.08643}{{\tt arXiv:2303.08643}}.
\bibitem[{{Duev} et~al.(2021){Duev}, {Bolin}, {Graham}, {Kelley}, {Mahabal},
  {Bellm}, {Coughlin}, {Dekany}, {Helou}, {Kulkarni}, {Masci}, {Prince},
  {Riddle}, {Soumagnac} and {van der Walt}}]{Duev2021}
\bibinfo{author}{{Duev}, D.A.}, \bibinfo{author}{{Bolin}, B.T.},
  \bibinfo{author}{{Graham}, M.J.}, \bibinfo{author}{{Kelley}, M.S.P.},
  \bibinfo{author}{{Mahabal}, A.}, \bibinfo{author}{{Bellm}, E.C.},
  \bibinfo{author}{{Coughlin}, M.W.}, \bibinfo{author}{{Dekany}, R.},
  \bibinfo{author}{{Helou}, G.}, \bibinfo{author}{{Kulkarni}, S.R.},
  \bibinfo{author}{{Masci}, F.J.}, \bibinfo{author}{{Prince}, T.A.},
  \bibinfo{author}{{Riddle}, R.}, \bibinfo{author}{{Soumagnac}, M.T.},
  \bibinfo{author}{{van der Walt}, S.J.}, \bibinfo{year}{2021}.
\newblock \bibinfo{title}{{Tails: Chasing Comets with the Zwicky Transient
  Facility and Deep Learning}}.
\newblock \bibinfo{journal}{\aj} \bibinfo{volume}{161}, \bibinfo{pages}{218}.
\newblock \DOIprefix\doi{10.3847/1538-3881/abea7b},
  \href{http://arxiv.org/abs/2102.13352}{{\tt arXiv:2102.13352}}.
\bibitem[{{Duev} et~al.(2020){Duev}, {Duev} and {Bolin}}]{Duev2020MPECT2}
\bibinfo{author}{{Duev}, D.A.}, \bibinfo{author}{{Duev}, I.D.},
  \bibinfo{author}{{Bolin}, B.T.}, \bibinfo{year}{2020}.
\newblock \bibinfo{title}{{COMET C/2020 T2 (Palomar)}}.
\newblock \bibinfo{journal}{Minor Planet Electronic Circulars}
  \bibinfo{volume}{2020-U170}.
\bibitem[{{Duev} et~al.(2019){Duev}, {Mahabal}, {Ye}, {Tirumala}, {Belicki},
  {Dekany}, {Frederick}, {Graham}, {Laher}, {Masci}, {Prince}, {Riddle},
  {Rosnet} and {Soumagnac}}]{Duev2019}
\bibinfo{author}{{Duev}, D.A.}, \bibinfo{author}{{Mahabal}, A.},
  \bibinfo{author}{{Ye}, Q.}, \bibinfo{author}{{Tirumala}, K.},
  \bibinfo{author}{{Belicki}, J.}, \bibinfo{author}{{Dekany}, R.},
  \bibinfo{author}{{Frederick}, S.}, \bibinfo{author}{{Graham}, M.J.},
  \bibinfo{author}{{Laher}, R.R.}, \bibinfo{author}{{Masci}, F.J.},
  \bibinfo{author}{{Prince}, T.A.}, \bibinfo{author}{{Riddle}, R.},
  \bibinfo{author}{{Rosnet}, P.}, \bibinfo{author}{{Soumagnac}, M.T.},
  \bibinfo{year}{2019}.
\newblock \bibinfo{title}{{DeepStreaks: identifying fast-moving objects in the
  Zwicky Transient Facility data with deep learning}}.
\newblock \bibinfo{journal}{\mnras} \bibinfo{volume}{486},
  \bibinfo{pages}{4158--4165}.
\newblock \DOIprefix\doi{10.1093/mnras/stz1096},
  \href{http://arxiv.org/abs/1904.05920}{{\tt arXiv:1904.05920}}.
\bibitem[{{Fang} et~al.(2017){Fang}, {Li} and {Cohn}}]{Fang2017AL}
\bibinfo{author}{{Fang}, M.}, \bibinfo{author}{{Li}, Y.},
  \bibinfo{author}{{Cohn}, T.}, \bibinfo{year}{2017}.
\newblock \bibinfo{title}{{Learning how to Active Learn: A Deep Reinforcement
  Learning Approach}}, in: \bibinfo{booktitle}{Proceedings of the 2017
  Conference on Empirical Methods in Natural Language Processing},
  \bibinfo{publisher}{Association for Computational Linguistics},
  \bibinfo{address}{Copenhagen, Denmark}. pp. \bibinfo{pages}{595--605}.
\bibitem[{{Ferellec} et~al.(2023){Ferellec}, {Snodgrass}, {Fitzsimmons},
  {Ro{\.z}ek}, {Gardener}, {Smith}, {Medeiros}, {Opitom} and
  {Hsieh}}]{Ferellec2023}
\bibinfo{author}{{Ferellec}, L.}, \bibinfo{author}{{Snodgrass}, C.},
  \bibinfo{author}{{Fitzsimmons}, A.}, \bibinfo{author}{{Ro{\.z}ek}, A.},
  \bibinfo{author}{{Gardener}, D.}, \bibinfo{author}{{Smith}, R.},
  \bibinfo{author}{{Medeiros}, H.}, \bibinfo{author}{{Opitom}, C.},
  \bibinfo{author}{{Hsieh}, H.H.}, \bibinfo{year}{2023}.
\newblock \bibinfo{title}{{A targeted search for Main Belt Comets}}.
\newblock \bibinfo{journal}{\mnras} \bibinfo{volume}{518},
  \bibinfo{pages}{2373--2384}.
\newblock \DOIprefix\doi{10.1093/mnras/stac3199},
  \href{http://arxiv.org/abs/2211.01435}{{\tt arXiv:2211.01435}}.
\bibitem[{{Festou} et~al.(1993){Festou}, {Rickman} and {West}}]{Festou1993}
\bibinfo{author}{{Festou}, M.C.}, \bibinfo{author}{{Rickman}, H.},
  \bibinfo{author}{{West}, R.M.}, \bibinfo{year}{1993}.
\newblock \bibinfo{title}{{Comets}}.
\newblock \bibinfo{journal}{\aapr} \bibinfo{volume}{4},
  \bibinfo{pages}{363--447}.
\newblock \DOIprefix\doi{10.1007/BF00872944}.
\bibitem[{{Fraser} et~al.(2022){Fraser}, {Dones}, {Volk}, {Womack} and
  {Nesvorn{\'y}}}]{Fraser2022review}
\bibinfo{author}{{Fraser}, W.C.}, \bibinfo{author}{{Dones}, L.},
  \bibinfo{author}{{Volk}, K.}, \bibinfo{author}{{Womack}, M.},
  \bibinfo{author}{{Nesvorn{\'y}}, D.}, \bibinfo{year}{2022}.
\newblock \bibinfo{title}{{The Transition from the Kuiper Belt to the
  Jupiter-Family (Comets)}}.
\newblock \bibinfo{journal}{arXiv e-prints} ,
  \bibinfo{pages}{arXiv:2210.16354}\DOIprefix\doi{10.48550/arXiv.2210.16354},
  \href{http://arxiv.org/abs/2210.16354}{{\tt arXiv:2210.16354}}.
\bibitem[{{Gladman} et~al.(2008){Gladman}, {Marsden} and
  {Vanlaerhoven}}]{Gladman2008}
\bibinfo{author}{{Gladman}, B.}, \bibinfo{author}{{Marsden}, B.G.},
  \bibinfo{author}{{Vanlaerhoven}, C.}, \bibinfo{year}{2008}.
\newblock \bibinfo{title}{{Nomenclature in the Outer Solar System}}, in:
  \bibinfo{editor}{{Barucci}, M.A.}, \bibinfo{editor}{{Boehnhardt}, H.},
  \bibinfo{editor}{{Cruikshank}, D.P.}, \bibinfo{editor}{{Morbidelli}, A.},
  \bibinfo{editor}{{Dotson}, R.} (Eds.), \bibinfo{booktitle}{The Solar System
  Beyond Neptune}, pp. \bibinfo{pages}{43--57}.
\bibitem[{{Hausen} and {Robertson}(2020)}]{Hausen2020}
\bibinfo{author}{{Hausen}, R.}, \bibinfo{author}{{Robertson}, B.E.},
  \bibinfo{year}{2020}.
\newblock \bibinfo{title}{{Morpheus: A Deep Learning Framework for the
  Pixel-level Analysis of Astronomical Image Data}}.
\newblock \bibinfo{journal}{\apjs} \bibinfo{volume}{248}, \bibinfo{pages}{20}.
\newblock \DOIprefix\doi{10.3847/1538-4365/ab8868},
  \href{http://arxiv.org/abs/1906.11248}{{\tt arXiv:1906.11248}}.
\bibitem[{{Hill} et~al.(2013){Hill}, {Bolin}, {Kleyna}, {Denneau}, {Wainscoat},
  {Micheli}, {Armstrong}, {Molina} and {Sato}}]{Hill2013b}
\bibinfo{author}{{Hill}, R.E.}, \bibinfo{author}{{Bolin}, B.},
  \bibinfo{author}{{Kleyna}, J.}, \bibinfo{author}{{Denneau}, L.},
  \bibinfo{author}{{Wainscoat}, R.}, \bibinfo{author}{{Micheli}, M.},
  \bibinfo{author}{{Armstrong}, J.D.}, \bibinfo{author}{{Molina}, M.},
  \bibinfo{author}{{Sato}, H.}, \bibinfo{year}{2013}.
\newblock \bibinfo{title}{{Comet P/2013 R3 (Catalina-Panstarrs)}}.
\newblock \bibinfo{journal}{Central Bureau Electronic Telegrams}
  \bibinfo{volume}{3658}.
\bibitem[{{Howell}(2000)}]{Howell2000}
\bibinfo{author}{{Howell}, S.B.}, \bibinfo{year}{2000}.
\newblock \bibinfo{title}{{Handbook of CCD Astronomy}}.
\bibitem[{{Hsieh} et~al.(2015){Hsieh}, {Denneau}, {Wainscoat},
  {Sch{\"o}rghofer}, {Bolin}, {Fitzsimmons}, {Jedicke}, {Kleyna}, {Micheli},
  {Vere{\v s}}, {Kaiser}, {Chambers}, {Burgett}, {Flewelling}, {Hodapp},
  {Magnier}, {Morgan}, {Price}, {Tonry} and {Waters}}]{Hsieh2015}
\bibinfo{author}{{Hsieh}, H.H.}, \bibinfo{author}{{Denneau}, L.},
  \bibinfo{author}{{Wainscoat}, R.J.}, \bibinfo{author}{{Sch{\"o}rghofer}, N.},
  \bibinfo{author}{{Bolin}, B.}, \bibinfo{author}{{Fitzsimmons}, A.},
  \bibinfo{author}{{Jedicke}, R.}, \bibinfo{author}{{Kleyna}, J.},
  \bibinfo{author}{{Micheli}, M.}, \bibinfo{author}{{Vere{\v s}}, P.},
  \bibinfo{author}{{Kaiser}, N.}, \bibinfo{author}{{Chambers}, K.C.},
  \bibinfo{author}{{Burgett}, W.S.}, \bibinfo{author}{{Flewelling}, H.},
  \bibinfo{author}{{Hodapp}, K.W.}, \bibinfo{author}{{Magnier}, E.A.},
  \bibinfo{author}{{Morgan}, J.S.}, \bibinfo{author}{{Price}, P.A.},
  \bibinfo{author}{{Tonry}, J.L.}, \bibinfo{author}{{Waters}, C.},
  \bibinfo{year}{2015}.
\newblock \bibinfo{title}{{The main-belt comets: The Pan-STARRS1 perspective}}.
\newblock \bibinfo{journal}{\icarus} \bibinfo{volume}{248},
  \bibinfo{pages}{289--312}.
\newblock \DOIprefix\doi{10.1016/j.icarus.2014.10.031},
  \href{http://arxiv.org/abs/1410.5084}{{\tt arXiv:1410.5084}}.
\bibitem[{{Jedicke} et~al.(2016){Jedicke}, {Bolin}, {Granvik} and
  {Beshore}}]{Jedicke2016}
\bibinfo{author}{{Jedicke}, R.}, \bibinfo{author}{{Bolin}, B.},
  \bibinfo{author}{{Granvik}, M.}, \bibinfo{author}{{Beshore}, E.},
  \bibinfo{year}{2016}.
\newblock \bibinfo{title}{{A fast method for quantifying observational
  selection effects in asteroid surveys}}.
\newblock \bibinfo{journal}{\icarus} \bibinfo{volume}{266},
  \bibinfo{pages}{173--188}.
\newblock \DOIprefix\doi{10.1016/j.icarus.2015.10.021}.
\bibitem[{{Jedicke} et~al.(2015){Jedicke}, {Granvik}, {Micheli}, {Ryan},
  {Spahr} and {Yeomans}}]{Jedicke2015}
\bibinfo{author}{{Jedicke}, R.}, \bibinfo{author}{{Granvik}, M.},
  \bibinfo{author}{{Micheli}, M.}, \bibinfo{author}{{Ryan}, E.},
  \bibinfo{author}{{Spahr}, T.}, \bibinfo{author}{{Yeomans}, D.K.},
  \bibinfo{year}{2015}.
\newblock \bibinfo{title}{{Surveys, Astrometric Follow-Up, and Population
  Statistics}}.
\newblock \bibinfo{journal}{Asteroids IV} , \bibinfo{pages}{795--813}.
\bibitem[{{Jedicke} et~al.(2002){Jedicke}, {Larsen} and {Spahr}}]{Jedicke2002}
\bibinfo{author}{{Jedicke}, R.}, \bibinfo{author}{{Larsen}, J.},
  \bibinfo{author}{{Spahr}, T.}, \bibinfo{year}{2002}.
\newblock \bibinfo{title}{{Observational Selection Effects in Asteroid
  Surveys}}.
\newblock \bibinfo{journal}{Asteroids III} , \bibinfo{pages}{71--87}.
\bibitem[{{Kelley} et~al.(2019){Kelley}, {Bodewits}, {Ye}, {Laher}, {Masci},
  {Monkewitz}, {Riddle}, {Rusholme}, {Shupe} and
  {Soumagnac}}]{Kelley2019Zchecker}
\bibinfo{author}{{Kelley}, M.S.P.}, \bibinfo{author}{{Bodewits}, D.},
  \bibinfo{author}{{Ye}, Q.}, \bibinfo{author}{{Laher}, R.R.},
  \bibinfo{author}{{Masci}, F.J.}, \bibinfo{author}{{Monkewitz}, S.},
  \bibinfo{author}{{Riddle}, R.}, \bibinfo{author}{{Rusholme}, B.},
  \bibinfo{author}{{Shupe}, D.L.}, \bibinfo{author}{{Soumagnac}, M.T.},
  \bibinfo{year}{2019}.
\newblock \bibinfo{title}{{ZChecker: Finding Cometary Outbursts with the Zwicky
  Transient Facility}}, in: \bibinfo{editor}{{Teuben}, P.J.},
  \bibinfo{editor}{{Pound}, M.W.}, \bibinfo{editor}{{Thomas}, B.A.},
  \bibinfo{editor}{{Warner}, E.M.} (Eds.), \bibinfo{booktitle}{Astronomical
  Data Analysis Software and Systems XXVII}, p. \bibinfo{pages}{471}.
\bibitem[{Kolen and Kremer(2001)}]{Kolen2001}
\bibinfo{author}{Kolen, J.F.}, \bibinfo{author}{Kremer, S.C.},
  \bibinfo{year}{2001}.
\newblock \bibinfo{title}{Gradient Flow in Recurrent Nets: The Difficulty of
  Learning LongTerm Dependencies}.
\newblock pp. \bibinfo{pages}{237--243}.
\newblock \DOIprefix\doi{10.1109/9780470544037.ch14}.
\bibitem[{Kubica et~al.(2005)Kubica, Moore, Connolly and Jedicke}]{Kubica2005}
\bibinfo{author}{Kubica, J.}, \bibinfo{author}{Moore, A.},
  \bibinfo{author}{Connolly, A.}, \bibinfo{author}{Jedicke, R.},
  \bibinfo{year}{2005}.
\newblock \bibinfo{title}{Efficiently identifying close track/observation pairs
  in continuous timed data}, \bibinfo{publisher}{SPIE}.
\bibitem[{{Mainzer} et~al.(2011){Mainzer}, {Bauer}, {Grav}, {Masiero}, {Cutri},
  {Dailey}, {Eisenhardt}, {McMillan}, {Wright}, {Walker}, {Jedicke}, {Spahr},
  {Tholen}, {Alles}, {Beck}, {Brandenburg}, {Conrow}, {Evans}, {Fowler},
  {Jarrett}, {Marsh}, {Masci}, {McCallon}, {Wheelock}, {Wittman}, {Wyatt},
  {DeBaun}, {Elliott}, {Elsbury}, {Gautier}, {Gomillion}, {Leisawitz},
  {Maleszewski}, {Micheli} and {Wilkins}}]{Mainzer2011a}
\bibinfo{author}{{Mainzer}, A.}, \bibinfo{author}{{Bauer}, J.},
  \bibinfo{author}{{Grav}, T.}, \bibinfo{author}{{Masiero}, J.},
  \bibinfo{author}{{Cutri}, R.M.}, \bibinfo{author}{{Dailey}, J.},
  \bibinfo{author}{{Eisenhardt}, P.}, \bibinfo{author}{{McMillan}, R.S.},
  \bibinfo{author}{{Wright}, E.}, \bibinfo{author}{{Walker}, R.},
  \bibinfo{author}{{Jedicke}, R.}, \bibinfo{author}{{Spahr}, T.},
  \bibinfo{author}{{Tholen}, D.}, \bibinfo{author}{{Alles}, R.},
  \bibinfo{author}{{Beck}, R.}, \bibinfo{author}{{Brandenburg}, H.},
  \bibinfo{author}{{Conrow}, T.}, \bibinfo{author}{{Evans}, T.},
  \bibinfo{author}{{Fowler}, J.}, \bibinfo{author}{{Jarrett}, T.},
  \bibinfo{author}{{Marsh}, K.}, \bibinfo{author}{{Masci}, F.},
  \bibinfo{author}{{McCallon}, H.}, \bibinfo{author}{{Wheelock}, S.},
  \bibinfo{author}{{Wittman}, M.}, \bibinfo{author}{{Wyatt}, P.},
  \bibinfo{author}{{DeBaun}, E.}, \bibinfo{author}{{Elliott}, G.},
  \bibinfo{author}{{Elsbury}, D.}, \bibinfo{author}{{Gautier}, IV, T.},
  \bibinfo{author}{{Gomillion}, S.}, \bibinfo{author}{{Leisawitz}, D.},
  \bibinfo{author}{{Maleszewski}, C.}, \bibinfo{author}{{Micheli}, M.},
  \bibinfo{author}{{Wilkins}, A.}, \bibinfo{year}{2011}.
\newblock \bibinfo{title}{{Preliminary Results from NEOWISE: An Enhancement to
  the Wide-field Infrared Survey Explorer for Solar System Science}}.
\newblock \bibinfo{journal}{ApJ} \bibinfo{volume}{731}, \bibinfo{pages}{53}.
\newblock \DOIprefix\doi{10.1088/0004-637X/731/1/53},
  \href{http://arxiv.org/abs/1102.1996}{{\tt arXiv:1102.1996}}.
\bibitem[{{Masci} et~al.(2019){Masci}, {Laher}, {Rusholme}, {Shupe}, {Groom},
  {Surace}, {Jackson}, {Monkewitz}, {Beck}, {Flynn}, {Terek}, {Landry},
  {Hacopians}, {Desai}, {Howell}, {Brooke}, {Imel}, {Wachter}, {Ye}, {Lin},
  {Cenko}, {Cunningham}, {Rebbapragada}, {Bue}, {Miller}, {Mahabal}, {Bellm},
  {Patterson}, {Juri{\'c}}, {Golkhou}, {Ofek}, {Walters}, {Graham}, {Kasliwal},
  {Dekany}, {Kupfer}, {Burdge}, {Cannella}, {Barlow}, {Van Sistine}, {Giomi},
  {Fremling}, {Blagorodnova}, {Levitan}, {Riddle}, {Smith}, {Helou}, {Prince}
  and {Kulkarni}}]{Masci2019}
\bibinfo{author}{{Masci}, F.J.}, \bibinfo{author}{{Laher}, R.R.},
  \bibinfo{author}{{Rusholme}, B.}, \bibinfo{author}{{Shupe}, D.L.},
  \bibinfo{author}{{Groom}, S.}, \bibinfo{author}{{Surace}, J.},
  \bibinfo{author}{{Jackson}, E.}, \bibinfo{author}{{Monkewitz}, S.},
  \bibinfo{author}{{Beck}, R.}, \bibinfo{author}{{Flynn}, D.},
  \bibinfo{author}{{Terek}, S.}, \bibinfo{author}{{Landry}, W.},
  \bibinfo{author}{{Hacopians}, E.}, \bibinfo{author}{{Desai}, V.},
  \bibinfo{author}{{Howell}, J.}, \bibinfo{author}{{Brooke}, T.},
  \bibinfo{author}{{Imel}, D.}, \bibinfo{author}{{Wachter}, S.},
  \bibinfo{author}{{Ye}, Q.Z.}, \bibinfo{author}{{Lin}, H.W.},
  \bibinfo{author}{{Cenko}, S.B.}, \bibinfo{author}{{Cunningham}, V.},
  \bibinfo{author}{{Rebbapragada}, U.}, \bibinfo{author}{{Bue}, B.},
  \bibinfo{author}{{Miller}, A.A.}, \bibinfo{author}{{Mahabal}, A.},
  \bibinfo{author}{{Bellm}, E.C.}, \bibinfo{author}{{Patterson}, M.T.},
  \bibinfo{author}{{Juri{\'c}}, M.}, \bibinfo{author}{{Golkhou}, V.Z.},
  \bibinfo{author}{{Ofek}, E.O.}, \bibinfo{author}{{Walters}, R.},
  \bibinfo{author}{{Graham}, M.}, \bibinfo{author}{{Kasliwal}, M.M.},
  \bibinfo{author}{{Dekany}, R.G.}, \bibinfo{author}{{Kupfer}, T.},
  \bibinfo{author}{{Burdge}, K.}, \bibinfo{author}{{Cannella}, C.B.},
  \bibinfo{author}{{Barlow}, T.}, \bibinfo{author}{{Van Sistine}, A.},
  \bibinfo{author}{{Giomi}, M.}, \bibinfo{author}{{Fremling}, C.},
  \bibinfo{author}{{Blagorodnova}, N.}, \bibinfo{author}{{Levitan}, D.},
  \bibinfo{author}{{Riddle}, R.}, \bibinfo{author}{{Smith}, R.M.},
  \bibinfo{author}{{Helou}, G.}, \bibinfo{author}{{Prince}, T.A.},
  \bibinfo{author}{{Kulkarni}, S.R.}, \bibinfo{year}{2019}.
\newblock \bibinfo{title}{{The Zwicky Transient Facility: Data Processing,
  Products, and Archive}}.
\newblock \bibinfo{journal}{\pasp} \bibinfo{volume}{131},
  \bibinfo{pages}{018003}.
\newblock \DOIprefix\doi{10.1088/1538-3873/aae8ac},
  \href{http://arxiv.org/abs/1902.01872}{{\tt arXiv:1902.01872}}.
\bibitem[{{Masiero} et~al.(2009){Masiero}, {Jedicke}, {{\v D}urech}, {Gwyn},
  {Denneau} and {Larsen}}]{Masiero2009}
\bibinfo{author}{{Masiero}, J.}, \bibinfo{author}{{Jedicke}, R.},
  \bibinfo{author}{{{\v D}urech}, J.}, \bibinfo{author}{{Gwyn}, S.},
  \bibinfo{author}{{Denneau}, L.}, \bibinfo{author}{{Larsen}, J.},
  \bibinfo{year}{2009}.
\newblock \bibinfo{title}{{The Thousand Asteroid Light Curve Survey}}.
\newblock \bibinfo{journal}{\icarus} \bibinfo{volume}{204},
  \bibinfo{pages}{145--171}.
\newblock \DOIprefix\doi{10.1016/j.icarus.2009.06.012},
  \href{http://arxiv.org/abs/0906.3339}{{\tt arXiv:0906.3339}}.
\bibitem[{{Muinonen} et~al.(2010){Muinonen}, {Belskaya}, {Cellino},
  {Delb{\`o}}, {Levasseur-Regourd}, {Penttil{\"a}} and
  {Tedesco}}]{Muinonen2010}
\bibinfo{author}{{Muinonen}, K.}, \bibinfo{author}{{Belskaya}, I.N.},
  \bibinfo{author}{{Cellino}, A.}, \bibinfo{author}{{Delb{\`o}}, M.},
  \bibinfo{author}{{Levasseur-Regourd}, A.C.}, \bibinfo{author}{{Penttil{\"a}},
  A.}, \bibinfo{author}{{Tedesco}, E.F.}, \bibinfo{year}{2010}.
\newblock \bibinfo{title}{{A three-parameter magnitude phase function for
  asteroids}}.
\newblock \bibinfo{journal}{\icarus} \bibinfo{volume}{209},
  \bibinfo{pages}{542--555}.
\newblock \DOIprefix\doi{10.1016/j.icarus.2010.04.003}.
\bibitem[{{Nesvorn{\'y}} et~al.(2023){Nesvorn{\'y}}, {Deienno}, {Bottke},
  {Jedicke}, {Naidu}, {Chesley}, {Chodas}, {Granvik}, {Vokrouhlick{\'y}},
  {Bro{\v{z}}}, {Morbidelli}, {Christensen}, {Shelly} and
  {Bolin}}]{Nesvorny2023NEO}
\bibinfo{author}{{Nesvorn{\'y}}, D.}, \bibinfo{author}{{Deienno}, R.},
  \bibinfo{author}{{Bottke}, W.F.}, \bibinfo{author}{{Jedicke}, R.},
  \bibinfo{author}{{Naidu}, S.}, \bibinfo{author}{{Chesley}, S.R.},
  \bibinfo{author}{{Chodas}, P.W.}, \bibinfo{author}{{Granvik}, M.},
  \bibinfo{author}{{Vokrouhlick{\'y}}, D.}, \bibinfo{author}{{Bro{\v{z}}}, M.},
  \bibinfo{author}{{Morbidelli}, A.}, \bibinfo{author}{{Christensen}, E.},
  \bibinfo{author}{{Shelly}, F.C.}, \bibinfo{author}{{Bolin}, B.T.},
  \bibinfo{year}{2023}.
\newblock \bibinfo{title}{{NEOMOD: A New Orbital Distribution Model for
  Near-Earth Objects}}.
\newblock \bibinfo{journal}{\aj} \bibinfo{volume}{166}, \bibinfo{pages}{55}.
\newblock \DOIprefix\doi{10.3847/1538-3881/ace040},
  \href{http://arxiv.org/abs/2306.09521}{{\tt arXiv:2306.09521}}.
\bibitem[{{Purdum} et~al.(2021){Purdum}, {Lin}, {Bolin}, {Sharma}, {Choi},
  {Bhalerao}, {Hanu{\v{s}}}, {Kumar}, {Quimby}, {van Roestel}, {Zhai},
  {Fernandez}, {Lisse}, {Bodewits}, {Fremling}, {Ryan Golovich}, {Hsu}, {Ip},
  {Ngeow}, {Saini}, {Shao}, {Yao}, {Ahumada}, {Anand}, {Andreoni}, {Burdge},
  {Burruss}, {Chang}, {Copperwheat}, {Coughlin}, {De}, {Dekany}, {Delacroix},
  {Drake}, {Duev}, {Graham}, {Hale}, {Kool}, {Kasliwal}, {Kostadinova},
  {Kulkarni}, {Laher}, {Mahabal}, {Masci}, {Mr{\'o}z}, {Neill}, {Riddle},
  {Rodriguez}, {Smith}, {Walters}, {Yan} and {Zolkower}}]{Purdum2021}
\bibinfo{author}{{Purdum}, J.N.}, \bibinfo{author}{{Lin}, Z.Y.},
  \bibinfo{author}{{Bolin}, B.T.}, \bibinfo{author}{{Sharma}, K.},
  \bibinfo{author}{{Choi}, P.I.}, \bibinfo{author}{{Bhalerao}, V.},
  \bibinfo{author}{{Hanu{\v{s}}}, J.}, \bibinfo{author}{{Kumar}, H.},
  \bibinfo{author}{{Quimby}, R.}, \bibinfo{author}{{van Roestel}, J.C.},
  \bibinfo{author}{{Zhai}, C.}, \bibinfo{author}{{Fernandez}, Y.R.},
  \bibinfo{author}{{Lisse}, C.M.}, \bibinfo{author}{{Bodewits}, D.},
  \bibinfo{author}{{Fremling}, C.}, \bibinfo{author}{{Ryan Golovich}, N.},
  \bibinfo{author}{{Hsu}, C.Y.}, \bibinfo{author}{{Ip}, W.H.},
  \bibinfo{author}{{Ngeow}, C.C.}, \bibinfo{author}{{Saini}, N.S.},
  \bibinfo{author}{{Shao}, M.}, \bibinfo{author}{{Yao}, Y.},
  \bibinfo{author}{{Ahumada}, T.}, \bibinfo{author}{{Anand}, S.},
  \bibinfo{author}{{Andreoni}, I.}, \bibinfo{author}{{Burdge}, K.B.},
  \bibinfo{author}{{Burruss}, R.}, \bibinfo{author}{{Chang}, C.K.},
  \bibinfo{author}{{Copperwheat}, C.M.}, \bibinfo{author}{{Coughlin}, M.},
  \bibinfo{author}{{De}, K.}, \bibinfo{author}{{Dekany}, R.},
  \bibinfo{author}{{Delacroix}, A.}, \bibinfo{author}{{Drake}, A.},
  \bibinfo{author}{{Duev}, D.}, \bibinfo{author}{{Graham}, M.},
  \bibinfo{author}{{Hale}, D.}, \bibinfo{author}{{Kool}, E.C.},
  \bibinfo{author}{{Kasliwal}, M.M.}, \bibinfo{author}{{Kostadinova}, I.S.},
  \bibinfo{author}{{Kulkarni}, S.R.}, \bibinfo{author}{{Laher}, R.R.},
  \bibinfo{author}{{Mahabal}, A.}, \bibinfo{author}{{Masci}, F.J.},
  \bibinfo{author}{{Mr{\'o}z}, P.J.}, \bibinfo{author}{{Neill}, J.D.},
  \bibinfo{author}{{Riddle}, R.}, \bibinfo{author}{{Rodriguez}, H.},
  \bibinfo{author}{{Smith}, R.M.}, \bibinfo{author}{{Walters}, R.},
  \bibinfo{author}{{Yan}, L.}, \bibinfo{author}{{Zolkower}, J.},
  \bibinfo{year}{2021}.
\newblock \bibinfo{title}{{Time-series and Phase-curve Photometry of the
  Episodically Active Asteroid (6478) Gault in a Quiescent State Using APO,
  GROWTH, P200, and ZTF}}.
\newblock \bibinfo{journal}{\apjl} \bibinfo{volume}{911}, \bibinfo{pages}{L35}.
\newblock \DOIprefix\doi{10.3847/2041-8213/abf2ca},
  \href{http://arxiv.org/abs/2102.13017}{{\tt arXiv:2102.13017}}.
\bibitem[{{Rehemtulla} et~al.(2023){Rehemtulla}, {Miller}, {Coughlin} and
  {Jegou du Laz}}]{Rehemtulla2023}
\bibinfo{author}{{Rehemtulla}, N.}, \bibinfo{author}{{Miller}, A.A.},
  \bibinfo{author}{{Coughlin}, M.W.}, \bibinfo{author}{{Jegou du Laz}, T.},
  \bibinfo{year}{2023}.
\newblock \bibinfo{title}{{$\texttt{BTSbot}$: A Multi-input Convolutional
  Neural Network to Automate and Expedite Bright Transient Identification for
  the Zwicky Transient Facility}}.
\newblock \bibinfo{journal}{arXiv e-prints} ,
  \bibinfo{pages}{arXiv:2307.07618}\DOIprefix\doi{10.48550/arXiv.2307.07618},
  \href{http://arxiv.org/abs/2307.07618}{{\tt arXiv:2307.07618}}.
\bibitem[{{Ro{\.z}ek} et~al.(2023){Ro{\.z}ek}, {Snodgrass}, {J{\o}rgensen},
  {Pravec}, {Bonavita}, {Rabus}, {Khalouei}, {Longa-Pe{\~n}a}, {Burgdorf},
  {Donaldson}, {Gardener}, {Crake}, {Sajadian}, {Bozza}, {Skottfelt},
  {Dominik}, {Fynbo}, {Hinse}, {Hundertmark}, {Rahvar}, {Southworth},
  {Tregloan-Reed}, {Kretlow}, {Rota}, {Peixinho}, {Andersen}, {Amadio},
  {Barrios-L{\'o}pez} and {Castillo Baeza}}]{Rozek2023}
\bibinfo{author}{{Ro{\.z}ek}, A.}, \bibinfo{author}{{Snodgrass}, C.},
  \bibinfo{author}{{J{\o}rgensen}, U.G.}, \bibinfo{author}{{Pravec}, P.},
  \bibinfo{author}{{Bonavita}, M.}, \bibinfo{author}{{Rabus}, M.},
  \bibinfo{author}{{Khalouei}, E.}, \bibinfo{author}{{Longa-Pe{\~n}a}, P.},
  \bibinfo{author}{{Burgdorf}, M.J.}, \bibinfo{author}{{Donaldson}, A.},
  \bibinfo{author}{{Gardener}, D.}, \bibinfo{author}{{Crake}, D.},
  \bibinfo{author}{{Sajadian}, S.}, \bibinfo{author}{{Bozza}, V.},
  \bibinfo{author}{{Skottfelt}, J.}, \bibinfo{author}{{Dominik}, M.},
  \bibinfo{author}{{Fynbo}, J.}, \bibinfo{author}{{Hinse}, T.C.},
  \bibinfo{author}{{Hundertmark}, M.}, \bibinfo{author}{{Rahvar}, S.},
  \bibinfo{author}{{Southworth}, J.}, \bibinfo{author}{{Tregloan-Reed}, J.},
  \bibinfo{author}{{Kretlow}, M.}, \bibinfo{author}{{Rota}, P.},
  \bibinfo{author}{{Peixinho}, N.}, \bibinfo{author}{{Andersen}, M.},
  \bibinfo{author}{{Amadio}, F.}, \bibinfo{author}{{Barrios-L{\'o}pez}, D.},
  \bibinfo{author}{{Castillo Baeza}, N.S.}, \bibinfo{year}{2023}.
\newblock \bibinfo{title}{{Optical Monitoring of the Didymos-Dimorphos Asteroid
  System with the Danish Telescope around the DART Mission Impact}}.
\newblock \bibinfo{journal}{\psj} \bibinfo{volume}{4}, \bibinfo{pages}{236}.
\newblock \DOIprefix\doi{10.3847/PSJ/ad0a64},
  \href{http://arxiv.org/abs/2311.01982}{{\tt arXiv:2311.01982}}.
\bibitem[{{Sato} et~al.(2022){Sato}, {Yoshimoto}, {Guido} and
  {Nakano}}]{Sato2022E3}
\bibinfo{author}{{Sato}, H.}, \bibinfo{author}{{Yoshimoto}, K.},
  \bibinfo{author}{{Guido}, E.}, \bibinfo{author}{{Nakano}, S.},
  \bibinfo{year}{2022}.
\newblock \bibinfo{title}{{COMET C/2022 E3}}.
\newblock \bibinfo{journal}{CBET} \bibinfo{volume}{5111}.
\bibitem[{{Schwamb} et~al.(2023){Schwamb}, {Jones}, {Yoachim}, {Volk},
  {Dorsey}, {Opitom}, {Greenstreet}, {Lister}, {Snodgrass}, {Bolin}, {Inno},
  {Bannister}, {Eggl}, {Solontoi}, {Kelley}, {Juri{\'c}}, {Lin}, {Ragozzine},
  {Bernardinelli}, {Chesley}, {Daylan}, {{\v{D}}urech}, {Fraser}, {Granvik},
  {Knight}, {Lisse}, {Malhotra}, {Oldroyd}, {Thirouin} and
  {Ye}}]{Schwamb2023ApJS}
\bibinfo{author}{{Schwamb}, M.E.}, \bibinfo{author}{{Jones}, R.L.},
  \bibinfo{author}{{Yoachim}, P.}, \bibinfo{author}{{Volk}, K.},
  \bibinfo{author}{{Dorsey}, R.C.}, \bibinfo{author}{{Opitom}, C.},
  \bibinfo{author}{{Greenstreet}, S.}, \bibinfo{author}{{Lister}, T.},
  \bibinfo{author}{{Snodgrass}, C.}, \bibinfo{author}{{Bolin}, B.T.},
  \bibinfo{author}{{Inno}, L.}, \bibinfo{author}{{Bannister}, M.T.},
  \bibinfo{author}{{Eggl}, S.}, \bibinfo{author}{{Solontoi}, M.},
  \bibinfo{author}{{Kelley}, M.S.P.}, \bibinfo{author}{{Juri{\'c}}, M.},
  \bibinfo{author}{{Lin}, H.W.}, \bibinfo{author}{{Ragozzine}, D.},
  \bibinfo{author}{{Bernardinelli}, P.H.}, \bibinfo{author}{{Chesley}, S.R.},
  \bibinfo{author}{{Daylan}, T.}, \bibinfo{author}{{{\v{D}}urech}, J.},
  \bibinfo{author}{{Fraser}, W.C.}, \bibinfo{author}{{Granvik}, M.},
  \bibinfo{author}{{Knight}, M.M.}, \bibinfo{author}{{Lisse}, C.M.},
  \bibinfo{author}{{Malhotra}, R.}, \bibinfo{author}{{Oldroyd}, W.J.},
  \bibinfo{author}{{Thirouin}, A.}, \bibinfo{author}{{Ye}, Q.},
  \bibinfo{year}{2023}.
\newblock \bibinfo{title}{{Tuning the Legacy Survey of Space and Time (LSST)
  Observing Strategy for Solar System Science}}.
\newblock \bibinfo{journal}{\apjs} \bibinfo{volume}{266}, \bibinfo{pages}{22}.
\newblock \DOIprefix\doi{10.3847/1538-4365/acc173},
  \href{http://arxiv.org/abs/2303.02355}{{\tt arXiv:2303.02355}}.
\bibitem[{{Scotti} et~al.(1992){Scotti}, {Gehrels} and
  {Rabinowitz}}]{Scotti1992}
\bibinfo{author}{{Scotti}, J.V.}, \bibinfo{author}{{Gehrels}, T.},
  \bibinfo{author}{{Rabinowitz}, D.L.}, \bibinfo{year}{1992}.
\newblock \bibinfo{title}{{Automated Detection of Asteroids in Real-Time with
  the Spacewatch Telescope}}, in: \bibinfo{editor}{{Harris}, A.W.},
  \bibinfo{editor}{{Bowell}, E.} (Eds.), \bibinfo{booktitle}{Asteroids, Comets,
  Meteors 1991}, p. \bibinfo{pages}{541}.
\bibitem[{{Sedaghat} et~al.(2024){Sedaghat}, {Chandler}, {Oldroyd}, {Trujillo},
  {Burris}, {Hsieh}, {Kueny}, {Farrell}, {DeSpain}, {Magbanua}, {Sheppard},
  {Mazzucato}, {Bosch}, {Shaw-Diaz}, {Gonano}, {Lamperti}, {da Silva Campos},
  {Goodwin}, {Terentev} and {Dukes}}]{Sedaghat2024}
\bibinfo{author}{{Sedaghat}, N.}, \bibinfo{author}{{Chandler}, C.O.},
  \bibinfo{author}{{Oldroyd}, W.J.}, \bibinfo{author}{{Trujillo}, C.A.},
  \bibinfo{author}{{Burris}, W.A.}, \bibinfo{author}{{Hsieh}, H.H.},
  \bibinfo{author}{{Kueny}, J.K.}, \bibinfo{author}{{Farrell}, K.A.},
  \bibinfo{author}{{DeSpain}, J.A.}, \bibinfo{author}{{Magbanua}, M.J.M.},
  \bibinfo{author}{{Sheppard}, S.S.}, \bibinfo{author}{{Mazzucato}, M.T.},
  \bibinfo{author}{{Bosch}, M.K.D.}, \bibinfo{author}{{Shaw-Diaz}, T.},
  \bibinfo{author}{{Gonano}, V.}, \bibinfo{author}{{Lamperti}, A.},
  \bibinfo{author}{{da Silva Campos}, J.A.}, \bibinfo{author}{{Goodwin}, B.L.},
  \bibinfo{author}{{Terentev}, I.A.}, \bibinfo{author}{{Dukes}, C.J.A.},
  \bibinfo{year}{2024}.
\newblock \bibinfo{title}{{2016 UU121: An Active Asteroid Discovery via
  AI-enhanced Citizen Science}}.
\newblock \bibinfo{journal}{Research Notes of the American Astronomical
  Society} \bibinfo{volume}{8}, \bibinfo{pages}{51}.
\newblock \DOIprefix\doi{10.3847/2515-5172/ad2b66}.
\bibitem[{{Sonnett} et~al.(2011){Sonnett}, {Kleyna}, {Jedicke} and
  {Masiero}}]{Sonnett2011}
\bibinfo{author}{{Sonnett}, S.}, \bibinfo{author}{{Kleyna}, J.},
  \bibinfo{author}{{Jedicke}, R.}, \bibinfo{author}{{Masiero}, J.},
  \bibinfo{year}{2011}.
\newblock \bibinfo{title}{{Limits on the size and orbit distribution of main
  belt comets}}.
\newblock \bibinfo{journal}{\icarus} \bibinfo{volume}{215},
  \bibinfo{pages}{534--546}.
\newblock \DOIprefix\doi{10.1016/j.icarus.2011.08.001},
  \href{http://arxiv.org/abs/1108.3095}{{\tt arXiv:1108.3095}}.
\bibitem[{{Tonry} et~al.(2018){Tonry}, {Denneau}, {Heinze}, {Stalder}, {Smith},
  {Smartt}, {Stubbs}, {Weiland} and {Rest}}]{Tonry2018}
\bibinfo{author}{{Tonry}, J.L.}, \bibinfo{author}{{Denneau}, L.},
  \bibinfo{author}{{Heinze}, A.N.}, \bibinfo{author}{{Stalder}, B.},
  \bibinfo{author}{{Smith}, K.W.}, \bibinfo{author}{{Smartt}, S.J.},
  \bibinfo{author}{{Stubbs}, C.W.}, \bibinfo{author}{{Weiland}, H.J.},
  \bibinfo{author}{{Rest}, A.}, \bibinfo{year}{2018}.
\newblock \bibinfo{title}{{ATLAS: A High-cadence All-sky Survey System}}.
\newblock \bibinfo{journal}{\pasp} \bibinfo{volume}{130},
  \bibinfo{pages}{064505}.
\newblock \DOIprefix\doi{10.1088/1538-3873/aabadf},
  \href{http://arxiv.org/abs/1802.00879}{{\tt arXiv:1802.00879}}.
\bibitem[{{Waszczak} et~al.(2013){Waszczak}, {Ofek}, {Aharonson}, {Kulkarni},
  {Polishook}, {Bauer}, {Levitan}, {Sesar}, {Laher}, {Surace} and {PTF
  Team}}]{Waszczak2013}
\bibinfo{author}{{Waszczak}, A.}, \bibinfo{author}{{Ofek}, E.O.},
  \bibinfo{author}{{Aharonson}, O.}, \bibinfo{author}{{Kulkarni}, S.R.},
  \bibinfo{author}{{Polishook}, D.}, \bibinfo{author}{{Bauer}, J.M.},
  \bibinfo{author}{{Levitan}, D.}, \bibinfo{author}{{Sesar}, B.},
  \bibinfo{author}{{Laher}, R.}, \bibinfo{author}{{Surace}, J.},
  \bibinfo{author}{{PTF Team}}, \bibinfo{year}{2013}.
\newblock \bibinfo{title}{{Main-belt comets in the Palomar Transient Factory
  survey - I. The search for extendedness}}.
\newblock \bibinfo{journal}{\mnras} \bibinfo{volume}{433},
  \bibinfo{pages}{3115--3132}.
\newblock \DOIprefix\doi{10.1093/mnras/stt951},
  \href{http://arxiv.org/abs/1305.7176}{{\tt arXiv:1305.7176}}.
\bibitem[{{Whidden} et~al.(2019){Whidden}, {Bryce Kalmbach}, {Connolly},
  {Jones}, {Smotherman}, {Bektesevic}, {Slater}, {Becker}, {Ivezi{\'c}},
  {Juri{\'c}}, {Bolin}, {Moeyens}, {F{\"o}rster} and {Golkhou}}]{Whidden2019}
\bibinfo{author}{{Whidden}, P.J.}, \bibinfo{author}{{Bryce Kalmbach}, J.},
  \bibinfo{author}{{Connolly}, A.J.}, \bibinfo{author}{{Jones}, R.L.},
  \bibinfo{author}{{Smotherman}, H.}, \bibinfo{author}{{Bektesevic}, D.},
  \bibinfo{author}{{Slater}, C.}, \bibinfo{author}{{Becker}, A.C.},
  \bibinfo{author}{{Ivezi{\'c}}, {\v{Z}}.}, \bibinfo{author}{{Juri{\'c}}, M.},
  \bibinfo{author}{{Bolin}, B.}, \bibinfo{author}{{Moeyens}, J.},
  \bibinfo{author}{{F{\"o}rster}, F.}, \bibinfo{author}{{Golkhou}, V.Z.},
  \bibinfo{year}{2019}.
\newblock \bibinfo{title}{{Fast Algorithms for Slow Moving Asteroids:
  Constraints on the Distribution of Kuiper Belt Objects}}.
\newblock \bibinfo{journal}{\aj} \bibinfo{volume}{157}, \bibinfo{pages}{119}.
\newblock \DOIprefix\doi{10.3847/1538-3881/aafd2d},
  \href{http://arxiv.org/abs/1901.02492}{{\tt arXiv:1901.02492}}.
\bibitem[{{Ye} et~al.(2019){Ye}, {Kelley}, {Bodewits}, {Bolin}, {Jones}, {Lin},
  {Bellm}, {Dekany}, {Duev}, {Groom}, {Helou}, {Kulkarni}, {Kupfer}, {Masci},
  {Prince} and {Soumagnac}}]{Ye2019Gault}
\bibinfo{author}{{Ye}, Q.}, \bibinfo{author}{{Kelley}, M.S.P.},
  \bibinfo{author}{{Bodewits}, D.}, \bibinfo{author}{{Bolin}, B.},
  \bibinfo{author}{{Jones}, L.}, \bibinfo{author}{{Lin}, Z.Y.},
  \bibinfo{author}{{Bellm}, E.C.}, \bibinfo{author}{{Dekany}, R.},
  \bibinfo{author}{{Duev}, D.A.}, \bibinfo{author}{{Groom}, S.},
  \bibinfo{author}{{Helou}, G.}, \bibinfo{author}{{Kulkarni}, S.R.},
  \bibinfo{author}{{Kupfer}, T.}, \bibinfo{author}{{Masci}, F.J.},
  \bibinfo{author}{{Prince}, T.A.}, \bibinfo{author}{{Soumagnac}, M.T.},
  \bibinfo{year}{2019}.
\newblock \bibinfo{title}{{Multiple Outbursts of Asteroid (6478) Gault}}.
\newblock \bibinfo{journal}{\apjl} \bibinfo{volume}{874}, \bibinfo{pages}{L16}.
\newblock \DOIprefix\doi{10.3847/2041-8213/ab0f3c},
  \href{http://arxiv.org/abs/1903.05320}{{\tt arXiv:1903.05320}}.

\end{thebibliography}





\end{document}